\documentclass[
superscriptaddress,
twocolumn,
amsmath,amssymb,
pra, aps,
nofootinbib
]{revtex4-2}    

\usepackage{graphicx}
\usepackage{dcolumn}
\usepackage{bm}
\usepackage{mathtools}
\usepackage{comment}
\usepackage{placeins}
\usepackage{braket}
\usepackage{hyperref}
\newcommand{\bb}[1]{\mathbf{#1}}

\begin{document}

\title{Simulating Quantum Light in Lossy Microring Resonators Driven by Strong Pulses}

\author{Youngbin Kim}
\affiliation{
School of Electrical Engineering, Korea Advanced Institute of Science and Technology (KAIST), Daejeon 34141, Republic of Korea}

\author{Seongjin Jeon}
\affiliation{
School of Electrical Engineering, Korea Advanced Institute of Science and Technology (KAIST), Daejeon 34141, Republic of Korea}

\author{Young-Ik Sohn}
\email{youngik.sohn@kaist.ac.kr}
\affiliation{
School of Electrical Engineering, Korea Advanced Institute of Science and Technology (KAIST), Daejeon 34141, Republic of Korea}

\date{\today}

\begin{abstract}
In this work, we present a quantum theory for pulsed photon pair generation in a single ring resonator. Our approach combines the Heisenberg picture input-output formalism with the Ikeda mapping from classical nonlinear optics. In doing so, we address the high-gain regime by incorporating non-perturbative effects, including self-phase modulation, cross-phase modulation, and time-ordering, which are roots for significantly different behaviors in the low-gain regime. We also account for optical losses by introducing an auxiliary waveguide, allowing for a more accurate representation of experimentally viable scenarios. Numerical simulations reveal that non-perturbative effects significantly distort transfer functions, making desirable operations challenging without careful optimization. We show that appropriate detuning of the pump frequency can mitigate these issues, leading to enhanced brightness and higher spectral purity in the high-gain regime. We further investigate the performance of a single ring resonator as a two-mode squeezer by analyzing various performance metrics under experimentally relevant optical loss conditions.
\end{abstract}

\maketitle

\section{INTRODUCTION}
Generating non-classical light is a pivotal task in fundamental quantum science and emerging quantum information technologies \cite{andersen201630,lounis2005single}. Among various methods for generating non-classical light, single- or two-mode squeezing based on nonlinear optical properties has been extensively used. By exposing a strong `pump' laser to a nonlinear material, it spontaneously emits non-classical, entangled photon pair referred to as `signal' and `idler'. When the spontaneous emission utilizes the material's $\chi_2$ nonlinearity, it is called spontaneous parametric down-conversion (SPDC), whereas utilizing $\chi_3$ nonlinearity is referred to as spontaneous four-wave mixing (SFWM). 

Nowadays, the integration of photon sources has become increasingly valuable as scalable quantum light sources, especially as the number of required optical components grows with the increasing complexity of quantum information processing tasks. For example, integrated quantum light sources are considered essential for realizing fault-tolerant quantum computation \cite{signorini2020chip,litinski2022active}. Integrated photon sources based on spontaneous emission are advantageous for scalable production of indistinguishable photons. Owing to these strengths, integrated SFWM and SPDC sources have been experimentally demonstrated in various applications, such as heralded single-photon generation \cite{paesani2020near}, squeezed vacuum generation \cite{stokowski2023integrated}, and Hong-Ou-Mandel experiment \cite{babel2023demonstration}. 

Between microring resonators and waveguides, the former have their strength in a very small footprint and large field enhancement. The coherent addition of multiple round-trips enhances the intra-cavity fields, making it easier to increase the photon pair generation rate with modest pump power. Additionally, the small mode volume resulting from integration facilitates greater nonlinearity, benefiting the brightness of the photon source \cite{luo2017chip}. High brightness is a desirable feature for various applications in both discrete-variable (DV) and continuous-variable (CV) regimes. Especially for CV quantum applications, simultaneously achieving high squeezing and a near-single temporal mode in an integrated device is a mandatory task \cite{vernon2019scalable}. 

Accordingly, the quantum theory of high-gain photon pair generation via ring resonator is an emerging area of study \cite{vendromin2024highly}. Under high-gain conditions, non-perturbative effects such as self-phase modulation (SPM), cross-phase modulation (XPM), and time-ordering effects become prominent. These effects have been extensively investigated in waveguides, demonstrating that in the high-gain regime, they alter the trends of generated mean photon numbers and the temporal mode structure compared to those in the low-gain regime \cite{quesada2020theory,bell2015effects,sinclair2016effect}. However, theoretical exploration of non-perturbative effects specifically in ring resonators remains limited. Moreover, although pulses play a crucial role in achieving single temporal mode operation, the impact of time-dependent non-perturbative effects on the temporal mode structure has not been thoroughly explored, to the best of our knowledge. To enable further high-gain applications of ring resonators in experiments, it is imperative to comprehensively explain and predict the nontrivial consequences of non-perturbative effects, losses, multi-photon contributions, and multi-mode features.

Over the past few decades, classical nonlinear dynamics such as second-harmonic generation and four-wave mixing in resonators have been actively investigated \cite{mckenna2022ultra,chen2012bistability}. Most studies have utilized temporal coupled-mode theory (TCMT), and its quantum optics version was developed by Vernon and Sipe \cite{vernon2015strongly}, where photon pair generation including non-perturbative effects was studied. However, the investigation has predominantly focused on continuous-wave (CW) pumping, leaving the effects of time-dependent non-perturbative phenomena on the modal structure of photon pair largely unexplored. Additionally, TCMT assumes a high-Q limit, where the wave circulates sufficiently long inside the resonator, assuming the field is uniformly distributed while it circulates inside the cavity. Consequently, this leads to deviations from the exact dynamics in the low-Q regime, such as the Lorentzian function inaccurately replaces the Airy function for resonance shape \cite{raymer2013quantum,jankowski2024ultrafast}.

These limitations can be overcome made by other approaches \cite{raymer2013quantum,alsing2017photon} that treat the intra-cavity field as a traveling wave. This formalism effectively captures the resonance function and commutation relations, but it has only considered in CW and low-gain regime, leaving pulsed pump operation unexplored. Meanwhile, Quesada \textit{et al.} \cite{quesada2020theory} explored the dynamics of photon pair generation in a waveguide driven by a broadband pump, achieving accurate simulation of transfer functions even in high-gain scenarios \cite{triginer2020understanding}. In this study, we extend the theory proposed by Alsing and Hach \cite{alsing2017photon} to encompass pulsed operations under high-gain conditions, by applying the methods from the work of Quesada \textit{et al.} \cite{quesada2020theory} combined with Ikeda mapping \cite{ikeda1980optical} from classical nonlinear optics.

In Sec. \ref{sec:theory}, the theory of photon pair generation in a ring resonator is explored. Beginning with a modeling of electric fields, the section presents the overall broadband input-output relation using transfer matrices. The preservation of the bosonic commutator property and the reduction to TCMT under high-Q limit are also discussed. In Sec. \ref{sec:numerical_simulation}, numerical simulations are conducted under low- and high-gain regime. We reveal that SPM and XPM together severely distort the cross-mode transfer function and hence influence the values of important performance metrics. Especially, we show that the ring resonator performs poorly if naively operated in the high-gain regime using the same input conditions as in the low-gain regime. To address the issue, detuned pump pulse is investigated to find optimal operating conditions. Based on numerical calculations, we thoroughly explore the behavior of the ring resonator with detuned pumping by calculating important measurable quantities. We further study the same system in the context of a resonator as a two-mode squeezer. The numerical model gives the prediction of squeezing and anti-squeezing parameters, as well as state and spectral purities under a few realistic optical loss scenarios.

Appendix \ref{app:sec:nonlinear_coefficients} provides the full expressions of nonlinear coefficients for SFWM, SPM, and XPM. In Appendix \ref{app:sec:low_gain_solution}, transfer functions are analytically derived in the low-gain regime. Appendix \ref{app:sec:physical_quantities} formulates the mean photon number, second-order correlation, and spectral purity using transfer matrix formalism. Appendix \ref{app:sec:squeezing_parameter} connects these formulations into the Gaussian optics framework, introducing how the detector can be incorporated and the squeezing parameters are being extracted. Lastly, Appendix \ref{app:sec:multiple_phantom_channels} extends the simulation framework presented in the main text to a more general loss model by incorporating multiple phantom channels, enabling a more precise treatment of propagation loss in low-Q ring resonators \cite{banic2022two}. 

\begin{figure}[htbp]
\includegraphics[width = \linewidth]{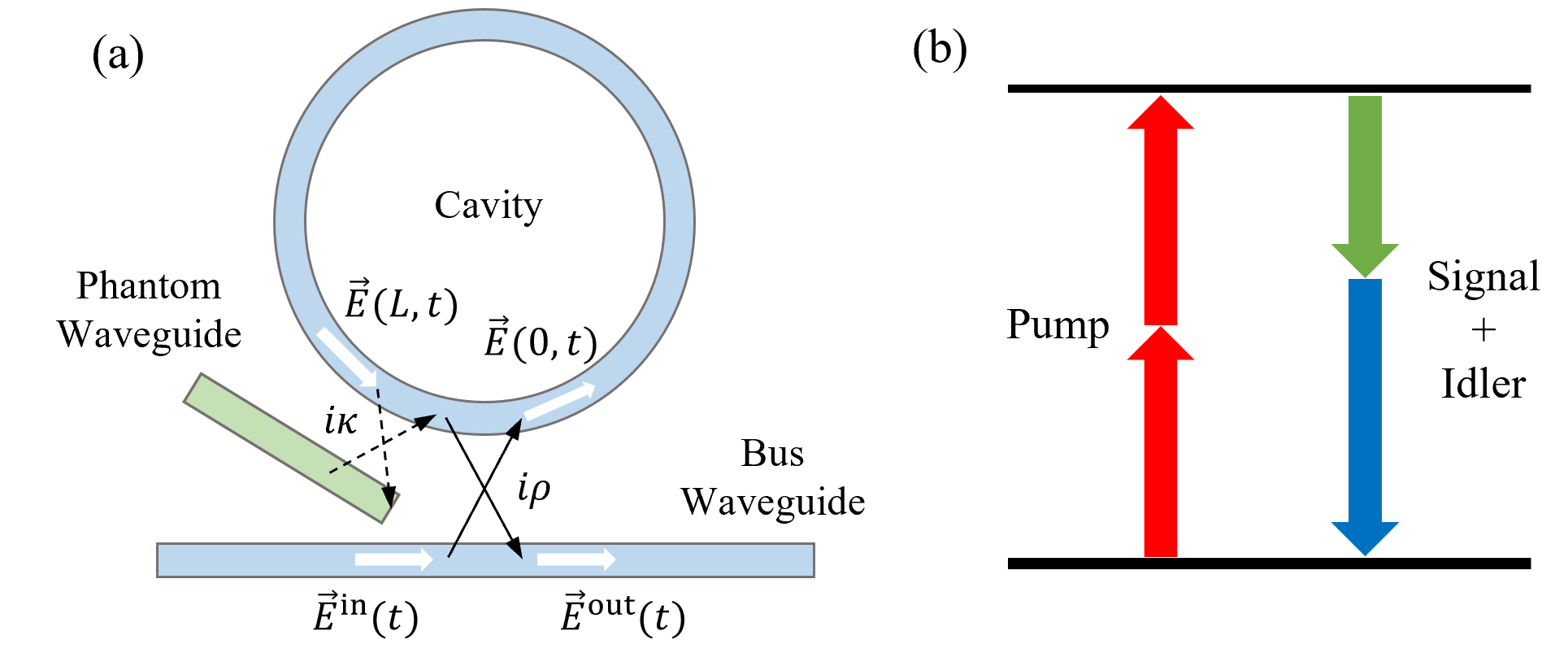}
\caption{(a) Schematic diagram of a microring resonator coupled to an external bus waveguide. A phantom waveguide is included to account for photon loss. (b) Illustration of single-pump SFWM.}\label{fig:fig_cav}
\end{figure}

\section{THEORY}
\label{sec:theory}
In this section we derive a transfer matrix for a single ring resonator incorporating nonlinear processes of SFWM, SPM, and XPM. The electric fields are modeled as traveling waves circulating via point-coupling between the ring resonator and the bus waveguide. We further consider a phantom waveguide to account for the loss of quantized modes \cite{vernon2015strongly}. The schematic diagram of a ring resonator coupled to a bus and a phantom waveguide, as well as the illustration of single-pump SFWM, is shown in Fig. \ref{fig:fig_cav}. 

While a single phantom channel is considered in this main text to minimize the computational tasks, we clarify that such configuration is accurate in the high-Q regime. To capture the impact of loss in low-Q resonators, it is necessary to extend to multiple phantom channels as discussed in Appendix \ref{app:sec:multiple_phantom_channels}. The framework developed in the Appendix is much more general in a sense that it is applicable for arbitrary coupling strengths between the resonator and a physical waveguide. With a straightforward extension from the single phantom channel model to a multiple phantom channel model, the proposed framework is robust across a wide range of Q-factors, including both high-Q and low-Q regimes. However, such treatment in high-Q regime only increases the computational complexity without gaining much improvement in accuracy. In the main text, we limit our focus on high-Q regime and adopt a single phantom channel configuration.

\subsection{Pump dynamics}
\label{sec:pump_dynamics}
In this subsection, we introduce the dynamics of pulsed pump in a ring resonator under the semi-classical and undepleted pump approximation. With these assumptions, the electric field of the pump mode can be expressed as Eq. (\ref{eq:pump_field}). Here, $\beta_p(z,\omega)$ denotes the spectral wavefunction at position $z$, $k_p(\omega)$ represents the effective propagation constant, and $\vec e_p(\vec r_\bot)$ signifies the normalized modal field of the pump mode. 
\begin{subequations}
\begin{gather}
\label{eq:pump_field}
\vec E_p(\vec r, t) = \cfrac{1}{\sqrt{2\pi}}\int d\omega \beta_p(z,\omega)e^{i(k_p(\omega)z-\omega t)}\vec e_p(\vec r_\bot) + \mathrm{c.c}, \\
\label{eq:pump_field_norm}
1 = \int dr_\bot (\vec e_p(\vec r_\bot)\times \vec h_p^*(\vec r_\bot) - \vec h_p(\vec r_\bot)\times\vec e_p^*(\vec r_\bot))\cdot \hat z 
\end{gather}
\end{subequations}
The Fourier-transform of $\beta_p(z,\omega)$ with reference to the simulation center frequency $\bar\omega_p$ is given by
\begin{equation}
\label{eq:pump_ft}
\tilde\beta_p(z,t) = \cfrac{1}{\sqrt{2\pi}}\int d\omega \beta_p(z,\omega) e^{-i(\omega-\bar\omega_p)t},
\end{equation}
where $\tilde\beta_p(z,t)$ represents the temporal wavefunction at position $z$ under a moving frame at the group velocity of pump, $v_p$, and a rotating frame of $\bar\omega_p$. The center frequency can be chosen freely; however in most cases, it is conveniently selected as the center frequency of the pump pulse or the pump resonance. By linearizing the effective propagation constant $k_p(\omega) =\bar k_p + (\omega-\bar\omega_p)/v_p$, where $\bar k_p = k_p(\bar \omega_p)$, the electric field of the pump can be expressed using the temporal wavefunction as shown in Eq. (\ref{eq:pump_field_temp}).
\begin{equation}
\label{eq:pump_field_temp}
\vec E_p(\vec r, t) = \tilde\beta_p\left(z,t-\cfrac{z}{v_p}\right)e^{i(\bar{k}_p z - \bar\omega_p t)}\vec e_p(\vec r_\bot) + \mathrm{c.c}.
\end{equation}
The temporal retardation of $t-z/v_p$ arises from the fact that $\tilde\beta_p(z,t)$ is measured in a moving frame, while $\vec E_p(\vec r, t)$ is measured in a fixed frame. The absolute square of the temporal wavefunction represents the power of the pump at a given time (i.e., $P_p(z,t) = |\tilde\beta_p(z,t)|^2$) with the modal field normalization as specified in Eq. (\ref{eq:pump_field_norm}). This electric field representation is applicable to both waveguide and intra-cavity fields. 

In this study, we consider a cavity coupled to an external bus waveguide at $z_{\mathrm{wg}}=0$ of the waveguide coordinate. The length of cavity is $L$, with coupling transmission and reflection coefficients denoted as $\tau$ and $\rho$ (for the pump, subscript $p$ as $\tau_p$ and $\rho_p$). We denote the pump field flowing into (out of) the cavity as $\tilde\beta_p^{\mathrm{in(out)}}(t)$ and the intra-cavity field at $(z,t)$ as $\tilde\beta_p(z, t-z/v_p)$. 

Due to the coupling between the resonator and the waveguide, the intra-cavity electric fields $\vec E_p(z=0, t)$ and $\vec E_p(z=L, t)$ experience discontinuity. The boundary condition can be modeled using a unitary matrix of a beam-splitter, expressed as
\begin{equation}
\label{eq:pump_bc}
\begin{pmatrix}
\tilde \beta_p(0,t) \\ \tilde\beta_p^{\mathrm{out}}(t)
\end{pmatrix} = 
\begin{pmatrix}
\tau_p & i\rho_p \\ i\rho_p & \tau_p
\end{pmatrix}
\begin{pmatrix}
\tilde \beta_p(L,t-T_p)e^{i\bar k_p L} \\ \tilde\beta_p^{\mathrm{in}}(t)
\end{pmatrix},
\end{equation}
where $T_p$ is the single round-trip time of the pump. With the phase relationship $e^{i\bar k_p L} = e^{i(\bar\omega_p -\omega_p^0)T_p}$ where $\omega_p^0$ is the nearest resonance frequency of the pump mode, the first row of Eq. (\ref{eq:pump_bc}) simplifies to:
\begin{equation}
\label{eq:pump_bc1}
\tilde\beta_p(0,t) = \tau_p\tilde\beta_p(L,t-T_p)e^{i(\bar\omega_p -\omega_p^0)T_p} + i\rho_p\tilde\beta_p^{\mathrm{in}}(t).
\end{equation}

Meanwhile, the round-trip traveling wave dynamics including SPM is given by
\begin{equation}
\label{eq:pump_dynamics}
\cfrac{\partial}{\partial z}\tilde\beta_p(z,t) = -\cfrac{\alpha_p}{2}\tilde\beta_p(z,t) + i\gamma_{\mathrm{spm}} e^{-\alpha_p z}|\tilde\beta_p(0,t)|^2\tilde\beta_p(z,t),
\end{equation}
where $\alpha_p$ and $\gamma_{\mathrm{spm}}$ are the propagation loss of the pump and the nonlinear coefficient of SPM, respectively \cite{agrawal2000nonlinear}. The dynamics described by Eq. (\ref{eq:pump_dynamics}) has an analytic solution:
\begin{equation}
\label{eq:pump_dynamics_sol}
\tilde\beta_p(L,t) = \tilde\beta_p(0,t)e^{-\alpha_pL/2}e^{i\gamma_{\mathrm{spm}}L_{\mathrm{eff}}|\tilde\beta_p(0,t)|^2},
\end{equation}
where $L_{\mathrm{eff}} =(1-e^{-\alpha_p L})/\alpha_p$ is the effective nonlinear interaction length. 

Combining the round-trip evolution Eq. (\ref{eq:pump_dynamics_sol}) and the boundary condition Eq. (\ref{eq:pump_bc1}), the overall dynamics of the pump is described by:
\begin{align}
\label{eq:ikeda_map}
\tilde\beta_p(0,t) = \tau_pe^{-\alpha_p L/2}&\tilde\beta_p(0,t-T_p)e^{i(\bar\omega_p-\omega_p^0)T_p}\times\\&e^{i\gamma_{\mathrm{spm}}L_{\mathrm{eff}}|\tilde\beta_p(0,t-T_p)|^2}+ i\rho_p\tilde\beta_p^{\mathrm{in}}(t). \nonumber
\end{align}
Given the flowing field $\tilde\beta_p^{\mathrm{in}}(t)$, the intra-cavity field at every round-trip time $T_p$ can be iteratively obtained. The dynamics described by Eq. (\ref{eq:ikeda_map}) is known as Ikeda mapping \cite{ikeda1980optical}, though the effects of group velocity dispersion are neglected. Using the discretized information of $\tilde\beta_p(0,t)$, the pump spectrum $\beta_p(0,\omega)$ can be obtained via inverse Fourier transform:
\begin{equation}
\label{eq:pump_inv_ft}
\beta_p(0,\omega) = \cfrac{1}{\sqrt{2\pi}}\int dt \tilde\beta_p(0,t) e^{i(\omega-\bar\omega_p)t}.
\end{equation}

\subsection{Photon pair dynamics}
Based on the intra-cavity pump spectrum $\beta_p(0,\omega)$, the transfer functions for round-trip photon pair generation in the $(z,\omega)$ domain can be obtained. The theory developed by Quesada \textit{et al.} \cite{quesada2020theory} derives the traveling wave photon pair generation dynamics using a Hamiltonian approach, which allows efficient simulation based on Trotterization. We utilize the dynamics proposed in this paper, and extend it to the resonant case by applying the boundary condition of our system.

The quantized electric field of the signal (idler) mode can be expressed as follows: 
\begin{subequations}
\begin{align}
\label{eq:photon_field}
\vec E_{s(i)}(\vec r, t) = \sqrt{\cfrac{\hbar\bar\omega_{s(i)}}{2\pi}}\int d\omega &\psi_{s(i)}(z,\omega)\times \\ &e^{i(k_{s(i)}(\omega)z-\omega t)}\vec e_{s(i)}(\vec r_\bot) + \mathrm{h.c}, \nonumber
\\ 
\label{eq:photon_field_norm}
1 = \int dr_\bot (\vec e_{s(i)}(\vec r_\bot)\times &\vec h_{s(i)}^*(\vec r_\bot) -\\ &\vec h_{s(i)}(\vec r_\bot)\times\vec e_{s(i)}^*(\vec r_\bot))\cdot \hat z,\nonumber
\end{align}
\end{subequations}
where $\psi_{s(i)}(z,\omega)$ is the photon annihilation operator for the signal (idler), normalized as:
\begin{equation}
\label{eq:photon_operator_norm}
n_{s(i)}(z) = \int d\omega \braket{\psi_{s(i)}^\dagger(z,\omega) \psi_{s(i)}(z,\omega)}.
\end{equation}
The operators are defined in the moving frame of their group velocities $v_{s(i)}$. Transforming to a frame with velocity $v_p$, new operators are defined as
\begin{equation}
\label{eq:ref_change}
a_{s(i)}(z,\omega) = \psi_{s(i)}(z,\omega)e^{i(\omega-\bar\omega_{s(i)})\left(\frac{1}{v_{s(i)}}-\frac{1}{v_p}\right)z},
\end{equation}
so that the phase of the temporal walk-off between the signal (idler) and pump is accumulated through the propagation.

At the coupling point, photons experience discontinuity due to the boundary condition. The boundary condition is modeled to accommodate propagation loss via the phantom channel approach, since incorporating propagation loss directly into the round-trip dynamics would violate the delta commutation relation of the bosonic system \cite{vernon2015spontaneous}. We model a phantom waveguide positioned immediately before the physical bus waveguide. The phantom waveguide is coupled to the cavity with transmission and reflection coefficients of $\gamma_{s(i)} = e^{-\alpha_{s(i)}L/2}$ and $\kappa_{s(i)}=\sqrt{1-e^{-\alpha_{s(i)}L}}$, respectively, such that the loss of a single round-trip is coupled out. Accordingly, the boundary condition for the signal mode is represented by
\begin{align}
\label{eq:photon_bc}
\begin{pmatrix}
a_s(0,\omega) \\ a_s^{\mathrm{out}}(\omega) \\ f_s^{\mathrm{out}}(\omega)  \end{pmatrix} =& \begin{pmatrix}
\tau_s\gamma_s & i\rho_s & i\tau_s\kappa_s \\
i\rho_s\gamma_s & \tau_s & -\rho_s\kappa_s \\
i\kappa_s & 0 & \gamma_s
\end{pmatrix} \times \\
& \begin{pmatrix}
a_s(L,\omega)e^{i[(\bar\omega_s-\omega_s^0)T_s+(\omega-\bar\omega_s)T_p]} \\ a_s^{\mathrm{in}}(\omega) \\ f_s^{\mathrm{in}}(\omega)  \end{pmatrix}, \nonumber
\end{align}
where $a_s^{\mathrm{in(out)}}(\omega)$ denotes the photon operator flowing into (out of) the bus waveguide, while $f_s^{\mathrm{in(out)}}(\omega)$ denotes the photon operator flowing into (out of) the phantom waveguide. The boundary condition Eq. (\ref{eq:photon_bc}) is also valid for the idler mode, with only a change in the subscript from $s$ to $i$. The boundary condition given in Eq. (\ref{eq:photon_bc}) can be simplified to the form:
\begin{subequations}
\label{eq:photon_bc_simple}
\begin{align}
\label{eq:photon_bc_simple1}
a_s(0,\omega) = \tau_s\gamma_sa_s(L,\omega)&e^{i[(\bar\omega_s-\omega_s^0)T_s+(\omega-\bar\omega_s)T_p]} \nonumber \\
+&i\rho_sa_s^{\mathrm{in}}(\omega) + i\tau_s\kappa_s f_s^{\mathrm{in}}(\omega), \\
\label{eq:photon_bc_simple2}
\tau_s a_s^{\mathrm{out}}(\omega) = a_s^{\mathrm{in}}(\omega) &+ i\rho_s a_s(0,\omega), \\
\label{eq:photon_bc_simple3}
\tau_s\gamma_s f_s^{\mathrm{out}}(\omega) = \tau_s f_s^{\mathrm{in}}(\omega) &+ i\kappa_s a_s(0,\omega) + \kappa_s\rho_s a_s^{\mathrm{in}}(\omega).
\end{align}
\end{subequations}

Inside the resonator, the dynamics of a single-round trip mediated by a broadband pump will mix the frequency components of the signal and idler photons, necessitating the discretization of frequency for broadband calculations. Accordingly, the boundary condition must be reformulated into a broadband matrix form to integrate with the round-trip dynamics. Henceforth, the operator vector will be defined by discretizing the frequency components into N points, given by $\omega_{s(i),n} = \omega_{s(i),0} + n\Delta\omega|_{n=0}^{N-1}$. The column vectors are defined as
\begin{equation}
\label{eq:operator_discretization}
\vec q_s = \begin{pmatrix} q_s(\omega_{s,0}) \\ q_s(\omega_{s,1}) \\... \\q_s(\omega_{s,N-1})
\end{pmatrix},
\vec q_i^\dagger = \begin{pmatrix} q_i^\dagger(\omega_{i,0}) \\ q_i^\dagger(\omega_{i,1}) \\... \\q_i^\dagger(\omega_{i,N-1})
\end{pmatrix},
\vec q =\begin{pmatrix} \vec q_s \\ \vec q_i^\dagger \end{pmatrix},
\end{equation}
where $q$ could be $a$ or $f$. This discretization approach is not limited to operators but should also be applied to the coefficients appeared in Eq. (\ref{eq:photon_bc_simple}). The coefficients corresponding to each frequency component are expanded to a diagonal square matrices. We define the matrices:
\begin{align}
\label{eq:t_r_mat}
T  = \begin{pmatrix} \tau_s I_N & O_N \\ O_N & \tau_i I_N\end{pmatrix}&, \
R = \begin{pmatrix} i\rho_s I_N & O_N \\ O_N & -i\rho_i I_N\end{pmatrix}, \\
\Gamma = \begin{pmatrix} \gamma_s I_N & O_N \\ O_N & \gamma_i I_N\end{pmatrix}&, \
K = \begin{pmatrix} i\kappa_s I_N & O_N \\ O_N & -i\kappa_i I_N\end{pmatrix}, \nonumber\\
E =& \begin{pmatrix} E_s & O_N \\ O_N & (E_i)^* \end{pmatrix} \nonumber \\
(E_{s(i)}^{n,m} =\delta_{n,m}e&^{i[(\bar\omega_{s(i)}-\omega^0_{s(i)})T_{s(i)}+(\omega_{s(i),n}-\bar\omega_{s(i)})T_p]}),\nonumber
\end{align}
where $I_N$ and $O_N$ are identity and zero matrices of size $N\times N$, respectively. The matrices $T$ and $R$ account for transmission and reflection to the bus waveguide, while $\Gamma$ and $K$ represent those interactions with the phantom waveguide. The block matrices $E_s$ and $E_i$ correspond to the accumulated phase in the electric fields of the signal and idler modes. In principle, the matrices $T$, $R$, $\Gamma$, and $K$ can describe frequency-dependent coupling by adjusting value of diagonal elements. However, in most cases, the resonance bandwidth of a ring resonator is sufficiently narrow so that frequency dependence becomes negligible, allowing the coefficients for each optical mode to be treated as frequency-independent. Using the definitions from Eq. (\ref{eq:operator_discretization}) and Eq. (\ref{eq:t_r_mat}), the boundary conditions in Eq. (\ref{eq:photon_bc_simple}) are transformed into the matrix form as follows:
\begin{subequations}
\begin{align}
\label{eq:photon_bc_mat1}
\vec a(0) &= T\Gamma E \vec a(L) + R \vec a^{\mathrm{in}} + TK\vec f^{\mathrm{in}}, \\
\label{eq:photon_bc_mat2}
T\vec a^{\mathrm{out}} &= \vec a^{\mathrm{in}} + R\vec a(0), \\
\label{eq:photon_bc_mat3}
T\Gamma \vec f^{\mathrm{out}} &= T\vec f^{\mathrm{in}} - KR \vec a^{\mathrm{in}} + K\vec a (0).
\end{align}
\end{subequations}

To solve Eq. (\ref{eq:photon_bc_mat1}), $\vec a(L)$ must be expressed in terms of $\vec a(0)$, which necessitates determining the round-trip transfer matrix. The frequency-domain dynamics of SFWM, including XPM, were derived by Quesada et al. \cite{quesada2020theory} and can be described as follows:
\begin{subequations}
\label{eq:photon_pair_dynamics}
\begin{align}
\label{eq:signal_dynamics}
\cfrac{\partial}{\partial z}a_s(z,\omega) &= i\Delta k_s(\omega)a_s(z,\omega)  \\
+& i\cfrac{2\gamma_{\mathrm{xpm},s}}{2\pi}\int d\omega' \mathcal{E}_p(\omega-\omega')a_s(z,\omega') \nonumber\\
+& i\cfrac{\gamma_{\mathrm{sfwm}}}{2\pi}\int d\omega' B_p(\omega+\omega')a_i^\dagger(z,\omega')e^{-i\left(\Delta\bar k - \frac{\Delta\bar\omega}{v_p}\right)z},\nonumber
\\
\label{eq:idler_dynamics}
\cfrac{\partial}{\partial z}a_i^\dagger(z,\omega) &= -i\Delta k_i(\omega)a_i^\dagger(z,\omega)  \\
-& i\cfrac{2\gamma_{\mathrm{xpm},i}}{2\pi}\int d\omega' \mathcal{E}_p^*(\omega-\omega')a_i^\dagger(z,\omega') \nonumber\\
-& i\cfrac{\gamma_{\mathrm{sfwm}}}{2\pi}\int d\omega' B_p^*(\omega+\omega')a_s(z,\omega')e^{i\left(\Delta\bar k - \frac{\Delta\bar\omega}{v_p}\right)z}.\nonumber
\end{align}
\end{subequations}
Here, $\Delta \bar k$ and $\Delta\bar\omega$ represent the phase mismatch and frequency mismatch at the simulation center frequency. Their definitions are:
\begin{subequations}
\begin{align}
\label{eq:phase_mismatch}
\Delta\bar k &= \bar k_s + \bar k_i - 2\bar k_p, \\
\label{eq:freq_mismatch}
\Delta\bar \omega &= \bar \omega_s + \bar \omega_i - 2\bar \omega_p.
\end{align}
\end{subequations}
Meanwhile, $\mathcal{E}_p$ and $B_p$ are the autocorrelation function and effective pump function, respectively, determined by $\beta_p(0,\omega)$ obtained from Eq. (\ref{eq:pump_inv_ft}). These functions are given by:
\begin{subequations}
\begin{align}
\label{eq:E_p}
\mathcal{E}_p(\omega-\omega') &= \int d\omega'' \beta_p^*(0,\omega''-(\omega-\omega'))\beta_p(0,\omega''),
\\
\label{eq:B_p}
B_p(\omega+\omega') &=  \int d\omega'' \beta_p(0, (\omega+\omega')-\omega'')\beta_p(0,\omega'').
\end{align}
\end{subequations}
Additionally, $\Delta k_{s(i)}(\omega) = (\omega-\bar\omega_{s(i)})\left(\cfrac{1}{v_{s(i)}}-\cfrac{1}{v_p}\right)$ describes the linear phase due to temporal walk-off between the pump and signal (idler). The nonlinear coefficients $\gamma_{\mathrm{xpm},s(i)}$ and $\gamma_{\mathrm{sfwm}}$ quantify the strength of XPM between the pump and signal (idler), and the strength of SFWM, respectively. Their full expressions are provided in Appendix \ref{app:sec:nonlinear_coefficients}.

The matrix form of the photon pair dynamics, whose continuous form is given Eq. (\ref{eq:photon_pair_dynamics}), can be expressed as below:
\begin{subequations}
\label{eq:photon_pair_dynamics_matrix}
\begin{align}
\cfrac{\partial}{\partial z}\vec a(z) &= i
\begin{pmatrix} G(z) & F(z) \\ -F^\dagger(z) & -H^\dagger(z) \end{pmatrix}\vec a(z), \\
F^{n,m}(z) =& \cfrac{\gamma_{\mathrm{sfwm}}}{2\pi}B_p(\omega_{s,n}+\omega_{i,m})e^{-i\left(\Delta\bar k - \frac{\Delta\bar\omega}{v_p}\right)z}\Delta\omega, \\
G^{n,m}(z) = & \Delta k_s(\omega_{s,n})\delta_{n,m} + \cfrac{2\gamma_{\mathrm{xpm},s}}{2\pi}\mathcal{E}_p(\omega_{s,n}-\omega_{s,m})\Delta\omega,\\
H^{n,m}(z) = &\Delta k_i(\omega_{i,n})\delta_{n,m} + \cfrac{2\gamma_{\mathrm{xpm},i}}{2\pi}\mathcal{E}_p^*(\omega_{i,n}-\omega_{i,m})\Delta\omega.
\end{align}
\end{subequations}
The matrix differential equation from Eq. (\ref{eq:photon_pair_dynamics_matrix}) can only be solved numerically. Let the solution be expressed as:
\begin{equation}
\label{eq:photon_dynamics_matrix_sol}
\vec a(L) = U\vec a(0),
\end{equation}
where $U$ is the round-trip transfer matrix.

By substituting Eq. (\ref{eq:photon_dynamics_matrix_sol}) into Eq. (\ref{eq:photon_bc_mat1}), $\vec a(0)$ can be expressed in terms of $\vec a^{\mathrm{in}}$ and $\vec f^{\mathrm{in}}$ as
\begin{equation}
\label{eq:photon_bc1_sol}
\vec a(0) = (I-T\Gamma E U)^{-1}R \vec a^{\mathrm{in}} +(I-T\Gamma EU)^{-1}TK\vec f^{\mathrm{in}}.
\end{equation}
With defining $Q$ matrix as:
\begin{equation}
\label{eq:Q_matrix}
Q \equiv I-T\Gamma E U,
\end{equation}
and substituting Eq. (\ref{eq:photon_bc1_sol}) into Eq. (\ref{eq:photon_bc_mat2}) and Eq. (\ref{eq:photon_bc_mat3}), we obtain the complete input-output relation for the signal and idler. Let the transfer matrix be denoted as $S$, with its components defined as:
\begin{subequations}
\label{eq:IO_relation}
\begin{align}
\begin{pmatrix} \vec a^{\mathrm{out}} \\ \vec f^{\mathrm{out}} \end{pmatrix} &= 
\overbrace{\begin{pmatrix} S^{aa} & S^{af} \\ S^{fa} & S^{ff} \end{pmatrix}}^{S}
\begin{pmatrix} \vec a^{\mathrm{in}} \\ \vec f^{\mathrm{in}} \end{pmatrix}, \\
S^{aa} &= T^{-1}(I+RQ^{-1}R), \\
S^{af} &= T^{-1}RQ^{-1}KT,  \\
S^{fa} &= T^{-1}\Gamma^{-1}K(Q^{-1}-I)R, \\
S^{ff} &= \Gamma^{-1} (I + T^{-1}KQ^{-1}KT).
\end{align}
\end{subequations}
The block matrix $S^{pq}$ ($p,q\in\{a, f\}$) is a transfer matrix that describes the contribution from the operators $\vec q^\text{in}$ to $\vec p^\text{out}$. Each block matrix can be decomposed into four square matrices of size $N\times N$. For instance, $S^{aa}$ can be expressed as follows:
\begin{equation}
\label{eq:S_aa_decompose}
S^{aa} = 
\begin{pmatrix} S^{aa}_{ss} & S^{aa}_{si} \\ 
(S^{aa}_{is})^* & (S^{aa}_{ii})^*
 \end{pmatrix}.
\end{equation}
The two block diagonal components, $S^{aa}_{ss}$ and $(S^{aa}_{ii})^*$, are same-mode transfer matrices for $a_s$ and $a_i^\dagger$, respectively, representing the contributions from $a_s$ to $a_s$ and from $a_i^\dagger$ to $a_i^\dagger$. Meanwhile, the two block off-diagonal components, $S^{aa}_{si}$ and $(S^{aa}_{is})^*$, indicate cross-mode contributions from $a_i^\dagger$ to $a_s$ and from $a_s$ to $a_i^\dagger$, which is a consequence of nonlinear mixing by SFWM. In the low-gain, lossless regime, the cross-mode transfer matrix $S^{aa}_{si}$ is approximately equal to the joint spectral amplitude (JSA) of photon pair \cite{ansari2018tailoring}. 

In the following, $\vec a_s^{\mathrm{out}}$ and $\vec a_i^{\mathrm{out},\dagger}$ are expressed in the discrete form as:
\begin{subequations}
\label{eq:io_matrix_form}
\begin{align}
\vec a_s^{\mathrm{out}} &= S^{aa}_{ss} \vec a_s^{\mathrm{in}} + S^{aa}_{si} \vec a_i^{\mathrm{in},\dagger}+S^{af}_{ss} \vec f_s^{\mathrm{in}} + S^{af}_{si} \vec f_i^{\mathrm{in},\dagger}, \\
\vec a_i^{\mathrm{out},\dagger} &= (S^{aa}_{is})^* \vec a_s^{\mathrm{in}} + (S^{aa}_{ii})^* \vec a_i^{\mathrm{in},\dagger}+\nonumber \\ &\quad\quad\quad\quad\quad\quad\quad\quad(S^{af}_{is})^* \vec f_s^{\mathrm{in}} + (S^{af}_{ii})^* \vec f_i^{\mathrm{in},\dagger}.
\end{align}
\end{subequations}
The input-output relation in continuous form is written as
\begin{subequations}
\label{eq:io_conti_form}
\begin{align}
&a_s^{\mathrm{out}}(\omega) \nonumber \\&= \int d\omega' [S^{aa}_{ss}(\omega,\omega') a_s^{\mathrm{in}}(\omega') +  S^{aa}_{si}(\omega,\omega') (a_i^{\mathrm{in}}(\omega'))^\dagger  \nonumber \\ 
& + S^{af}_{ss}(\omega,\omega') f_s^{\mathrm{in}}(\omega') +  S^{af}_{si}(\omega,\omega') (f_i^{\mathrm{in}}(\omega'))^\dagger], \\
&(a_i^{\mathrm{out}}(\omega))^\dagger \nonumber \\&= \int d\omega' [(S^{aa}_{is}(\omega,\omega'))^* a_s^{\mathrm{in}}(\omega') +  (S^{aa}_{ii}(\omega,\omega'))^* (a_i^{\mathrm{in}}(\omega'))^\dagger  \nonumber \\ 
&+ (S^{af}_{is}(\omega,\omega'))^* f_s^{\mathrm{in}}(\omega') +  (S^{af}_{ii}(\omega,\omega'))^* (f_i^{\mathrm{in}}(\omega'))^\dagger],
\end{align}
\end{subequations}
where the integration is defined from $-\infty$ to $+\infty$. Therefore, the relation between the discrete transfer matrix and the continuous transfer function is given by:
\begin{align}
[S^{pq}_{uv}]_{n,m} =  \Delta \omega &\cdot S^{pq}_{uv}(\omega_{u,n}, \omega_{v,m}) \nonumber\\
(p,q\in\{a, f\} &, \  u, v \in \{s, i\}). 
\end{align}

\subsection{Connection to TCMT}
To this point, we have derived a transfer function for photon pair in the frequency domain. Vernon and Sipe \cite{vernon2015strongly} previously introduced a framework for photon pair dynamics that led to the TCMT formalism, beginning with a rigorous Hamiltonian treatment. In this subsection, we validate our formalism by demonstrating its reduction to TCMT under the high-Q limit.

In the high-Q limit, the transmission and reflection of boundary conditions can be approximated as continuous decay processes, allowing the following expressions \cite{lugiato2018lugiato}:
\begin{subequations}
    \label{eq:high_Q_approx}
    \begin{align}
        \tau_m &= e^{-\kappa_{\mathrm{ex},m}T_m}\simeq 1, \\
        \gamma_m &=e^{-\kappa_{\mathrm{in},m}T_m}\simeq 1, \\
        \rho_m &\simeq \sqrt{2\kappa_{\mathrm{ex},m}T_m}, \\
        \kappa_m &\simeq \sqrt{2\kappa_{\mathrm{in},m}T_m}.
    \end{align}
\end{subequations}
Here, $\kappa_{\text{ex(in)},m}$ represents the external (internal) decay rate of mode $m$, arising from the coupling (propagation loss). 

Using Eq. (\ref{eq:high_Q_approx}), the transfer matrices $S^{aa}$ and $S^{af}$, as expressed in Eq. (\ref{eq:IO_relation}), reduces to
\begin{subequations}
    \begin{align}
        S^{aa} &\simeq I + RQ^{-1}R, \\
        S^{af} &\simeq RQ^{-1}K. 
    \end{align}
\end{subequations}
This simplifies the vectorized input-output relation, $\vec a^\text{out} = S^{aa}\vec a^\text{in} + S^{af}\vec f^\text{in}$, to
\begin{equation}
\label{eq:IO_a_cav}
    \vec a^\text{out}\simeq \vec a^\text{in} + R\underbrace{Q^{-1}\left(R\vec a^\text{in} + K\vec f^\text{in}\right)}_{\vec a^\text{cav}},
\end{equation}
indicating that $\vec a^\text{cav}$ corresponds to the intra-cavity operator $\vec a(0)$ in the high-Q limit. With this definition, Eq. (\ref{eq:IO_a_cav}) can be expressed as:
\begin{subequations}
\label{eq:langevin_vec}
    \begin{align}
    \label{eq:langevin_vec_1}
        \vec a^\text{out} &= \vec a^\text{in} + R\vec a^\text{cav}, \\
    \label{eq:langevin_vec_2}
        Q\vec a^\text{cav} &= R \vec a^\text{in} + K \vec f^\text{in}.
    \end{align}
\end{subequations}

In the high-Q limit, the mean-field approximation \cite{grelu2015nonlinear} also applies, where neglects a change of field during a single round-trip. Applying this approximation to Eq. (\ref{eq:photon_pair_dynamics_matrix}), the round-trip transfer matrix $U$ simplifies to
\begin{equation}
    U \simeq I + iL\begin{pmatrix}
        G(0) & F(0) \\ -F^\dagger(0) & -H^\dagger(0)
    \end{pmatrix}.
\end{equation}
The $Q$ matrix from Eq. (\ref{eq:Q_matrix}) then reduces to
\begin{equation}
        Q \rightarrow \begin{pmatrix}
            Q_{ss} & Q_{si}\\ Q_{si}^\dagger & (Q_{ii})^*
        \end{pmatrix} ,
\end{equation}
where the block matrices are defined as follows:
\begin{subequations}
\label{eq:Q_matrix_tcmt}
    \begin{align}
        Q_{ss}^{n,m} &= \left[\kappa_{\text{tot},s}-i(\omega_{s,n}-\omega_s^0)\right]T_s 
        \\
       &\quad\quad\quad -i\cfrac{2\gamma_{\text{xpm},s}L}{2\pi}\mathcal{E}_p(\omega_{s,n}-\omega_{s,m})\Delta\omega, \nonumber 
        \\
        Q_{ii}^{n,m} &= \left[\kappa_{\text{tot},i}-i(\omega_{i,n}-\omega_i^0)\right]T_i 
        \\
       &\quad\quad\quad -i\cfrac{2\gamma_{\text{xpm},i}L}{2\pi}\mathcal{E}_p(\omega_{i,n}-\omega_{i,m})\Delta\omega, \nonumber
        \\
        Q_{si}^{n,m} &= -i\cfrac{\gamma_\text{sfwm}L}{2\pi}B_p(\omega_{s,n} + \omega_{i,m})\Delta\omega.
    \end{align}
\end{subequations}

Substituting Eq. (\ref{eq:Q_matrix_tcmt}) into Eq. (\ref{eq:langevin_vec_2}), the continuous version of Eq. (\ref{eq:langevin_vec}) becomes:
\begin{subequations}
    \label{eq:langevin_ft}
    \begin{align}
    \label{eq:langevin_ft1}
         a^\text{out}_{s}(\omega) = a^\text{in}_{s}(\omega) + i&\sqrt{2\kappa_{\text{ex},s}T_{s}} a^\text{cav}_s(\omega),
        \\
    \label{eq:langevin_ft2}
        \left[\kappa_{\text{tot},s}-i(\omega-\omega_{s}^0)\right]T_s a_s^\text{cav}(\omega) &=
        \\
         i\cfrac{2\gamma_{\text{xpm},s}L}{2\pi}\int d\omega' &\mathcal{E}_p(\omega-\omega') a_s^\text{cav}(\omega') \nonumber
        \\
         +i\cfrac{\gamma_\text{sfwm}L}{2\pi}\int d\omega' &B_p(\omega+\omega')\left(a_i^\text{cav}(\omega')\right)^\dagger \nonumber
        \\
        +i\sqrt{2\kappa_{\text{ex},s}T_s}a_s^\text{in}&(\omega) + i\sqrt{2\kappa_{\text{in},s}T_s}f_s^\text{in}(\omega). \nonumber
    \end{align}
\end{subequations}
The corresponding equations for the idler mode can be obtained by exchanging the indices $s\leftrightarrow i$. These equations, given in Eq. (\ref{eq:langevin_ft}), represent the frequency-domain (Fourier-transformed) version of the TCMT framework, as originally developed by Vernon and Sipe \cite{vernon2015spontaneous}.

\subsection{Modal structure}
The transfer matrix $S$ can be decomposed as shown in Eq. (\ref{eq:sandwich}), where $U_1$ and $U_2$ are unitary matrices, and $C$ represents the transfer matrix of a lossless multi-mode optical parametric amplifier \cite{cui2021high}. The decomposition is given by:
\begin{subequations}
\label{eq:sandwich}
\begin{align}
S =& \ U_2 C U_1,
\\
U_2 =& \begin{pmatrix}
Rn^{-1}  &  K\Gamma^{-1}n^{-1} \\
K\Gamma^{-1}n^{-1} & -Rn^{-1}
\end{pmatrix},
\\
C =& \begin{pmatrix}\Gamma^{-1}T^{-1}-nT^{-1}Q^{-1}\Gamma n & O_{2N} \\ O_{2N} & I_{2N}  \end{pmatrix},
\\
U_1 =&\begin{pmatrix} -R\Gamma^{-1}n^{-1} & -KT\Gamma^{-1}n^{-1} \\ -KT\Gamma^{-1}n^{-1} & R\Gamma^{-1}n^{-1}
\end{pmatrix},
\\
n =& \sqrt{-\left(R^2 + K^2\Gamma^{-2}\right)}.
\end{align}
\end{subequations}
The unitary matrices $U_1$ and $U_2$ function as beam-splitters, facilitating the transformation between operators in the physical waveguide and those in the phantom waveguide. In the high-Q limit, these unitary matrices are reduced into the beam-splitters where the signal and idler modes experience reflectivities of $\kappa_{\mathrm{ex},s}/ \kappa_{\mathrm{tot},s}$ and $\kappa_{\mathrm{ex},i}/ \kappa_{\mathrm{tot},i}$, respectively, as illustrated in Eq. (\ref{eq:unitary_high_Q}).
\begin{align}
\label{eq:unitary_high_Q}
&  U_2 \stackrel{\mathrm{High-Q}}{=} -U_1\rightarrow \\
& \begin{pmatrix}
i\sqrt{\cfrac{\kappa_{\mathrm{ex},s}}{\kappa_{\mathrm{tot},s}}}I_N & O_N & i\sqrt{\cfrac{\kappa_{\mathrm{in},s}}{\kappa_{\mathrm{tot},s}}}I_N & O_N \\
O_N & -i\sqrt{\cfrac{\kappa_{\mathrm{ex},i}}{\kappa_{\mathrm{tot},i}}}I_N & O_N & -i\sqrt{\cfrac{\kappa_{\mathrm{in},i}}{\kappa_{\mathrm{tot},i}}}I_N \\
i\sqrt{\cfrac{\kappa_{\mathrm{in},s}}{\kappa_{\mathrm{tot},s}}}I_N & O_N & -i\sqrt{\cfrac{\kappa_{\mathrm{ex},s}}{\kappa_{\mathrm{tot},s}}}I_N & O_N \\
O_N & -i\sqrt{\cfrac{\kappa_{\mathrm{in},i}}{\kappa_{\mathrm{tot},i}}}I_N & O_N & i\sqrt{\cfrac{\kappa_{\mathrm{ex},i}}{\kappa_{\mathrm{tot},i}}}I_N
\end{pmatrix} \nonumber
\end{align}

Furthermore, the upper left block matrix of $C$, denoted as $C^{aa}$, represents the transfer matrix for lossless twin-beam generation, which characterizes two-mode squeezing. This matrix $C^{aa}$ can be decomposed as follows: 
\begin{equation}
\label{eq:C_aa_decomp}
C^{aa} = \begin{pmatrix}P_s & O_N \\ O_N & P_i^* \end{pmatrix}
\begin{pmatrix}  \cosh{(R^c)} & \sinh{(R^c)} \\ \sinh{(R^c)} & \cosh{(R^c)}\end{pmatrix}
\begin{pmatrix}Q_s & O_N \\ O_N & Q_i^* \end{pmatrix}^\dagger,
\end{equation}
where $P_j$ and $Q_j$ are unitary matrices, and $R^c$ is a non-negative, real diagonal matrix \cite{christ2013theory}. The $l^{\mathrm{th}}$ column of $P_{s(i)}$ corresponds to the $l^\mathrm{th}$ broadband output optical mode of signal (idler), while the $l^\mathrm{th}$ column of $Q_{s(i)}$ corresponds to the $l^\mathrm{th}$ broadband input optical mode of signal (idler). The diagonal elements in $R^c$, denoted as $\{r^c\}$, represent the two-mode squeezing parameters. Specifically, its $l^\text{th}$ component, $r_l^c$, corresponds to the squeezing parameter of the $l^\text{th}$ broadband mode. The broadband output modes represent the set of Schmidt mode of the overall system, which characterize the temporal mode structure of the resonator. The decomposition determines the set of Schmidt modes as well as the effective number of modes, quantified by the Schmidt number $\mathcal{K}$ or spectral purity $\mathcal{P}$, as defined in Appendix \ref{app:sec:physical_quantities}.

The two beam-splitters $U_1$ and $U_2$ mix the operators from the physical waveguide and the phantom waveguide, thereby limiting the maximum achievable squeezing due to the intrinsic losses of the cavity \cite{cui2021high}. Assuming that the signal and idler modes exhibit identical response to the cavity, the maximum achievable squeezing is constrained by the escape efficiency, $\eta_\text{esc}$, defined as:
\begin{equation}
\eta_\text{esc} = \cfrac{\kappa_\text{ex}}{\kappa_\text{in} + \kappa_\text{ex}}.
\end{equation}
Given the squeezing parameter $r^c$ from the two-mode squeezing matrix $C^{aa}$, the output squeezing parameter $r$ can be estimated using \cite{cui2021high, park2024single}:
\begin{equation}
\label{eq:eff_squeezing}
r = -\cfrac{1}{2}\ln\left(1-\eta_\text{esc}+\eta_\text{esc}e^{-2r^c}\right).
\end{equation}
Thus, the squeezing in decibels (dB) is bounded by $-10\log_{10}(1-\eta_\text{esc})$.

Meanwhile, the commutation relations for the output operators $\vec a^{\mathrm{out}}$ and $\vec f^{\mathrm{out}}$ must be preserved. The complete set of commutation relations are given by: 
\begin{subequations}
\label{eq:commutation}
\begin{align}
\left[a_{s(i)}^{\mathrm{out}}(\omega), (a_{s(i)}^{\mathrm{out}}(\omega'))^\dagger\right] =& \delta(\omega-\omega'),  \\
\left[f_{s(i)}^{\mathrm{out}}(\omega), (f_{s(i)}^{\mathrm{out}}(\omega'))^\dagger\right] =& \delta(\omega-\omega'), \\
\Big[a_s^{\mathrm{out}}(\omega), a_i^{\mathrm{out}}(\omega')\Big] = \Big[a_s^{\mathrm{out}}(\omega),& (a_i^{\mathrm{out}}(\omega'))^\dagger\Big] =0, \\
\Big[f_s^{\mathrm{out}}(\omega), f_i^{\mathrm{out}}(\omega')\Big] = \Big[f_s^{\mathrm{out}}(\omega),& (f_i^{\mathrm{out}}(\omega'))^\dagger\Big] =0, \\
\Big[a_{s(i)}^{\mathrm{out}}(\omega), f_{s(i)}^{\mathrm{out}}(\omega')\Big] = \Big[a_{s(i)}^{\mathrm{out}}(\omega),& (f_{s(i)}^{\mathrm{out}}(\omega'))^\dagger\Big] =0.
\end{align}
\end{subequations}
In compact matrix form, the commutation relation in Eq. (\ref{eq:commutation}) is expressed as \cite{weedbrook2012gaussian}:
\begin{equation}
\label{eq:commutation_mat}
S\mathbb{J}S^\dagger = \mathbb{J},
\end{equation}
where
\begin{equation}
\mathbb{J} = \begin{pmatrix}
I_N & O_N & O_N & O_N \\
O_N & -I_N & O_N & O_N \\
O_N & O_N & I_N & O_N \\
O_N & O_N & O_N & -I_N
\end{pmatrix}.
\end{equation}
The properties
\begin{equation}
U_1\mathbb{J}U_1^\dagger = U_2\mathbb{J}U_2^\dagger = C\mathbb{J}C^\dagger = \mathbb{J},
\end{equation}
ensure that the commutation relation in Eq. (\ref{eq:commutation_mat}) follows directly from Eq. (\ref{eq:sandwich}).

\section{NUMERICAL SIMULATION}
\label{sec:numerical_simulation}

In this section, we demonstrate a numerical simulation based on the dynamics described by Eq. (\ref{eq:ikeda_map}) and Eq. (\ref{eq:IO_relation}). For simplicity, we assume identical linear responses for the pump, signal, and idler modes within the ring resonator. The ring resonator has a radius of $200 \ \mathrm{\mu m}$, and the free-spectral range (FSR) for all three modes is $117 \ \mathrm{GHz}$. The propagation loss is set to $\alpha = 0.1 \ \mathrm{dB/cm}$, and the coupling coefficient between the ring resonator and the waveguide is $\rho = \sqrt{0.01}$. Given parameters were inferred from the previously reported SiN ring resonator experiment \cite{vaidya2020broadband,arrazola2021quantum}. 
Such configuration should be taken as an overcoupled regime as the external decay rate $\kappa_{\mathrm{ex}} = 588 \ \mathrm{MHz}$ exceeds the internal decay rate $\kappa_{\mathrm{in}} = 169 \ \mathrm{MHz}$. Corresponding escape efficiency is $\eta_\text{esc}=0.776$ for all three modes. Resonance wavelengths are chosen to be spaced by three times the FSR, which correspond to $\lambda_p^0 = 1554.2 \ \mathrm{nm}$, $\lambda_s^0 = 1551.4 \ \mathrm{nm}$, and $\lambda_i^0 = 1557.0 \ \mathrm{nm}$. At these resonance wavelengths, perfect phase matching and frequency matching are assumed (i.e., $\Delta\omega^0 = \Delta k^0 = 0$). The nonlinear coefficients $\gamma_{\mathrm{sfwm}}$, $\gamma_{\mathrm{spm}}$, $\gamma_{\mathrm{xpm},s}$, and $\gamma_{\mathrm{xpm},i}$ are all set to $1 \ \mathrm{W^{-1}m^{-1}}$. 

\subsection{Mode structure and photon statistics}

A transform-limited Gaussian pump pulse with a center wavelength of $1554.2 \ \mathrm{nm}$ (zero detuning from resonance) and an intensity full-width at half-maximum (FWHM) of $283 \ \mathrm{MHz}$ is injected through the bus waveguide. This FWHM value is comparable to the resonance FWHM of $241 \ \mathrm{MHz}$. The pump energy ranges from $1 \ \mathrm{pJ}$ up to $600 \ \mathrm{pJ}$, where high-gain effects can be obviously seen. 

\begin{figure}[htbp]
\includegraphics[width = \linewidth]{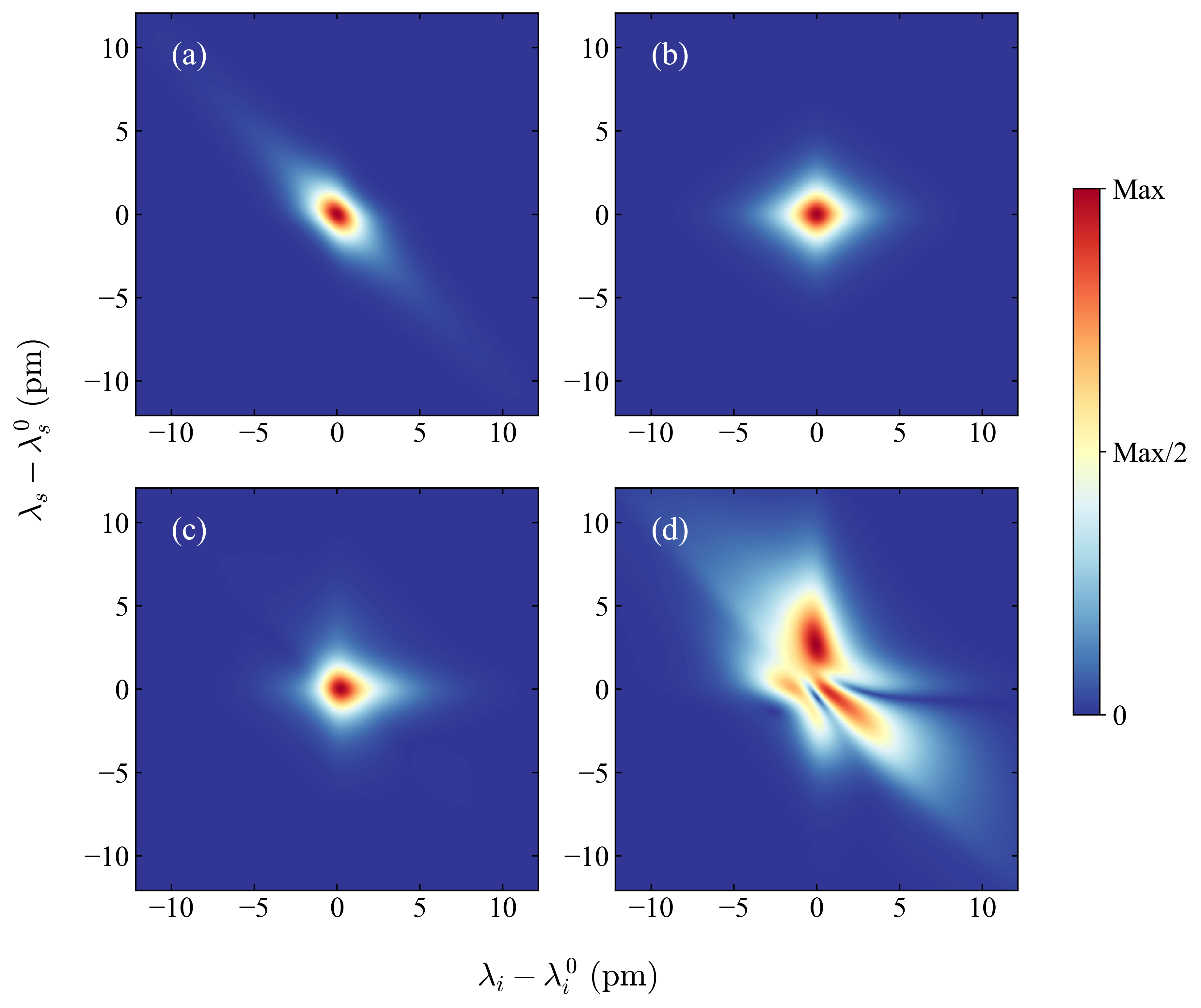}
\caption{Cross-mode transfer matrices at different scenarios: (a) $1 \ \mathrm{pJ}$, (b) $600 \ \mathrm{pJ}$ considering only SFWM, (c) $600 \ \mathrm{pJ}$ with SFWM and SPM, and (d) $600 \ \mathrm{pJ}$ with all effects incorporated.}\label{fig:fig_cm}
\end{figure}

The cross-mode transfer matrix $S^{aa}_{si}$ for various configurations are plotted in Fig. \ref{fig:fig_cm}. Panel (a) shows the scenario with a low pump energy of $1 \ \mathrm{pJ}$, while panels (b), (c), and (d) illustrate scenarios with a pump energy of $600 \ \mathrm{pJ}$. To analyze the effects of SFWM, SPM, and XPM sequentially, Fig. \ref{fig:fig_cm}(b) depicts the case with only SFWM, Fig. \ref{fig:fig_cm}(c) includes SPM in addition to SFWM, and Fig. \ref{fig:fig_cm}(d) incorporates all effects. At low pump energy, the cross-mode transfer function matches the first-order perturbation solution of Eq. (\ref{eq:signal_transfer_function_si_lowgain}). However, at high pump energy, the transfer matrices exhibit distinct characteristics.

When only SFWM is considered, as shown in Fig. \ref{fig:fig_cm}(b), the shape of the cross-mode transfer function becomes more rounded. In this case, the resonance frequencies of the signal and idler are not relevant due to the absence of frequency pulling. In the perturbative regime, the magnitude of the cross-mode transfer function, as derived in Eq. (\ref{eq:signal_transfer_function_si_lowgain}), is linearly dependent on the pump pulse energy. As will be utilized for the simplified SFWM model later in this subsection, an increase in pump energy results in a linear amplification of the squeezing parameter, $r^c$, for each mode in the perturbative regime \cite{triginer2020understanding, christ2013theory}. The decomposition Eq. (\ref{eq:C_aa_decomp}) shows that the hyperbolic sine of the squeezing parameters determines the weight of the corresponding Schmidt mode. In the low-gain regime, since the hyperbolic sine functions can be well approximated by linear functions, the uniform amplification of the squeezing parameters applies to all Schmidt modes, leaving the cross-mode transfer function unchanged. However, in the high-gain regime, large squeezing parameters require the use of the full expression of the hyperbolic functions. Consequently, the coefficient of the dominant Schmidt mode becomes exponentially larger than all the other modes. This dominance of the first Schmidt mode results in a more rounded cross-mode transfer function and, therefore, improved spectral purity \cite{triginer2020understanding}.

\begin{figure}[htbp]
\includegraphics[width = \linewidth]{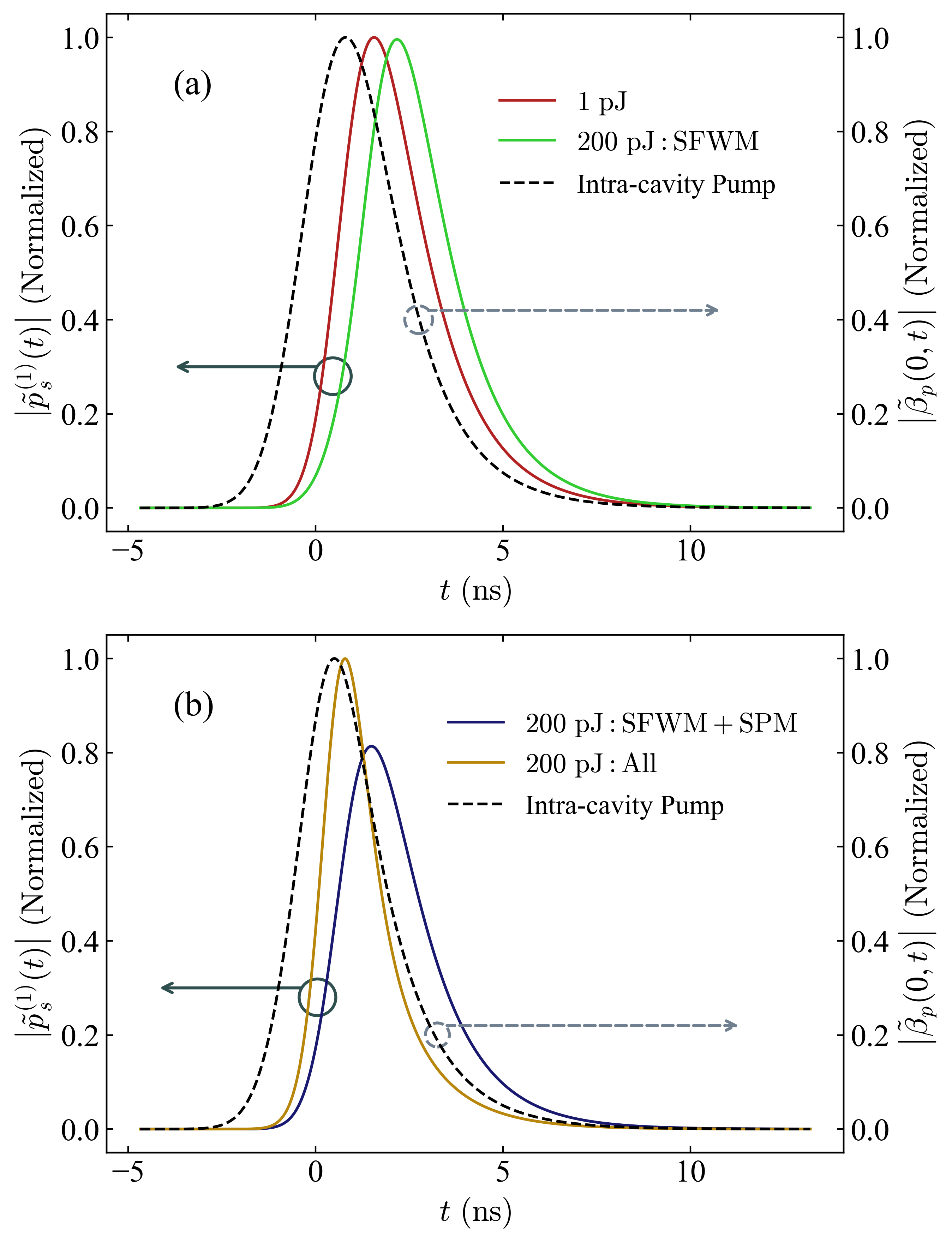}
\caption{Variation in the first temporal Schmidt mode of the signal mode, $\left|\tilde p_s^{(1)}(t)\right|$, under different conditions. The functions are normalized such that the maximum value among the two scenarios is unity. (a) The red line represents the case with a pump energy of $1 \ \mathrm{pJ}$, and the green line corresponds to a pump energy of $200 \ \mathrm{pJ}$, considering only SFWM. The dotted line indicates the normalized intra-cavity pump pulse in the absence of SPM. (b) The blue and yellow lines represent scenarios with a pump energy of $200 \ \mathrm{pJ}$, where SPM and XPM are progressively one after the other. Additionally, the dotted line represents the normalized intra-cavity pump pulse influenced by SPM.}\label{fig:fig_Sch}
\end{figure}

This exact treatment also accounts for the complete set of terms in the Magnus expansion \cite{quesada2014effects,vendromin2024highly}, thereby capturing time-ordering corrections. The impact of time-ordering is evident by the changes in the temporal mode structure \cite{quesada2022beyond,christ2013theory}. The red and green lines in Fig. \ref{fig:fig_Sch}(a) explicitly shows the variation in the first temporal Schmidt mode of the signal mode, $\tilde p^{(1)}_s(t)$, for pump energies of $1 \ \mathrm{pJ}$ and $200 \ \mathrm{pJ}$, considering only SFWM. The corresponding mean photon numbers for those two cases are around $1.4\times 10^{-4}$ and 71, respectively. The first temporal Schmidt mode is obtained by applying a Fourier transform to the first column of the matrix $P_s$, derived from the decomposition in Eq. (\ref{eq:C_aa_decomp}). Notably, significant variations in the temporal mode are observed even under the exclusive influence of SFWM. In the high-gain regime, the temporal mode's peak exhibits a delay compared to the low-gain scenario. This behavior can be attributed to the stimulated processes driven by early-converted photons, corresponding to the scattering dynamics of higher-order Magnus terms \cite{thekkadath2024gain}. The exact solution not only enables the observation of changes in the cross-mode transfer function but also provides insight into the system's temporal mode structure. 

\begin{figure}[t]
\includegraphics[width = \linewidth]{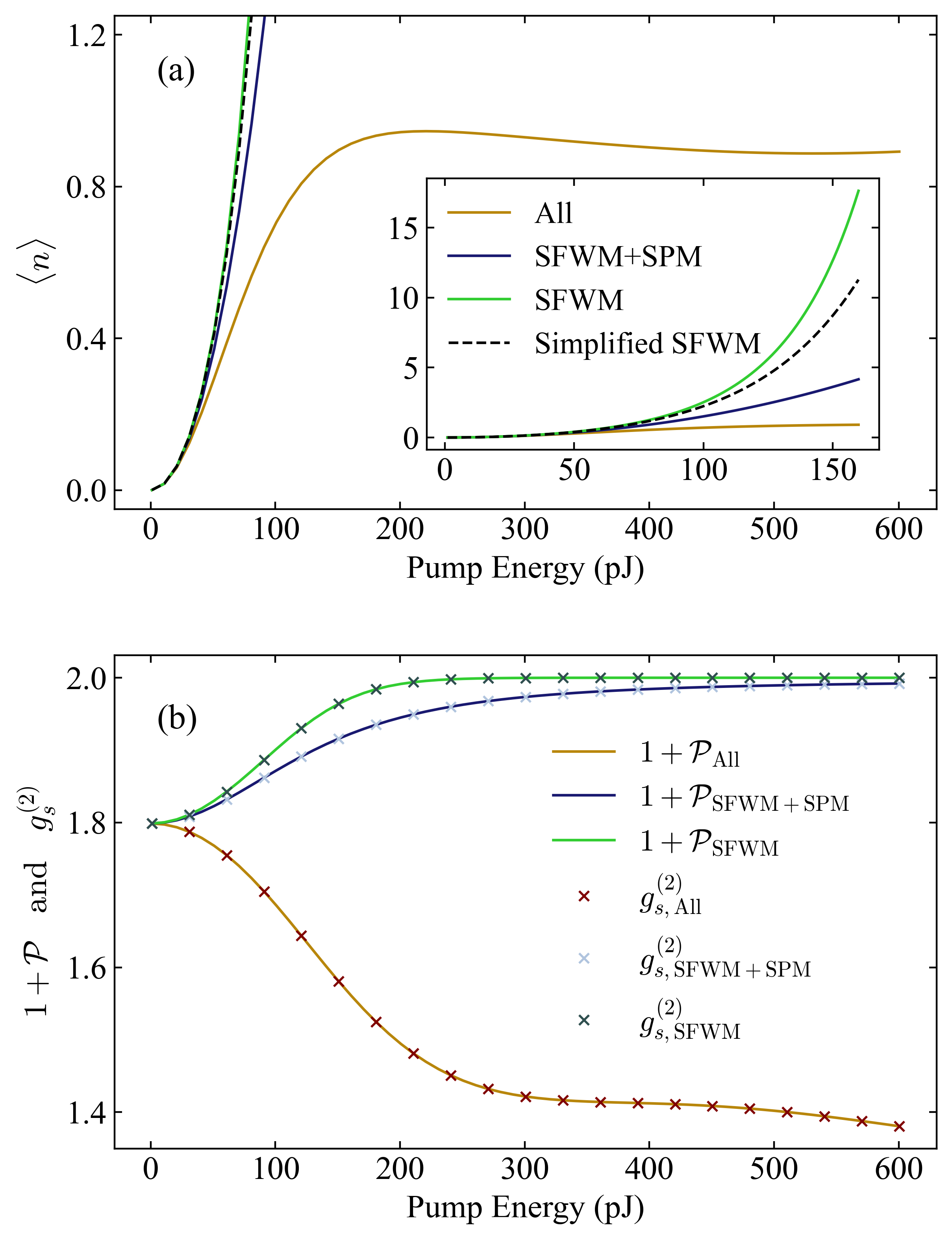}
\caption{(a) Mean photon number versus pump energy. The yellow line represents the scenario with all effects included, the blue line shows the scenario with SFWM and SPM included, and the green line corresponds to the scenario with only SFWM considered. The black dotted line indicates the simplified SFWM model, assuming that the squeezing parameters in $C^{aa}$ scales linearly with pump energy (i.e., $r^{c}\propto E_p$). (b) Spectral purity (solid lines) and second-order self-correlation (markers) as functions of pump energy.}\label{fig:fig_nP}
\end{figure}

Once we start considering SPM, still without XPM, the intra-cavity pump experiences variations by the influence of SPM (see two dotted lines in Fig. \ref{fig:fig_Sch}(a) and (b)). This modifies the effective pump function, $B_p(\omega+\omega')$, causing the transfer function to deviate from the case without SPM, as illustrated by the change from panel (b) to (c) of Fig. \ref{fig:fig_cm}. Despite this distortion, the central spectral island of the transfer function remains a single round shape because the signal and idler resonances are not influenced by XPM yet. Therefore, as shown by the blue line in Fig. \ref{fig:fig_nP}(b), the spectral purity can asymptotically approach unity including SPM, however, it is not true when XPM is included.

Under the influence of SPM, the circulating power within the ring is effectively reduced due to the resonance frequency pull in the pump mode. This results in a diminished strength of SFWM, as evident by the significantly slower increase in the mean photon number, shown by the blue line in the inset of Fig. \ref{fig:fig_nP}(a). 

The reduction in SFWM caused by SPM also impact the temporal mode structure. The reduced SFWM leads to weaker stimulated process mentioned earlier, and hence eventually mitigates the delay of the first temporal Schmidt mode. As a result, the blue line in panel (b) of Fig. \ref{fig:fig_Sch}, representing a pump energy of $200 \ \mathrm{pJ}$ with SFWM and SPM, and the red line of panel (a) exhibit very little difference in its temporal peak location. 

However, once XPM is included, the cross-mode transfer function undergoes significant distortion. The combination of time-dependent SPM and XPM not only shifts the resonance center of the three modes but also effectively distorts the resonance shape. Primarily due to XPM, the signal and idler resonances are pulled away far from the center, resulting in an unpredictably broader overall spectrum, as shown in Fig. \ref{fig:fig_cm}(d). Therefore, time-dependent XPM significantly impairs spectral purity, as illustrated by the yellow line in Fig. \ref{fig:fig_nP}(b), which shows a completely different trend compared to the green and blue lines.

Meanwhile, XPM brings the peak of the temporal Schmidt mode to an earlier time, as shown in the yellow line in Fig. \ref{fig:fig_Sch}(b), compared to the red line in panel (a). XPM pulls away the resonance frequencies of the signal and idler modes from its original values, a amount twice that of the pump mode shift influenced by SPM \cite{vernon2015strongly}. This induces a mismatch in energy conservation (see Eq. (\ref{eq:freq_mismatch})) between the resonance frequencies of the three modes. Consequently, the generated photons are effectively detuned from the shifted resonances, causing them to escape the resonator more rapidly and advancing the temporal peak. 

Based on this sequential analysis of nonlinear effects, to fully characterize the temporal mode structure in the high-gain regime, it is essential to consider all three effects: SFWM, SPM, and XPM.

Next, we investigate the scaling of the mean photon number. This value can be extracted from the transfer matrices, with detailed expressions provided in Appendix \ref{app:sec:physical_quantities}. The mean photon number is plotted in Fig. \ref{fig:fig_nP}(a). The yellow line represents the scenario where all effects are considered, the blue line corresponds to the case where XPM is excluded, and the green line illustrates the situation where only SFWM is taken into account. The black dashed line illustrates the simplified SFWM model, which assumes that the squeezing parameters in $C^{aa}$ scale linearly with pump energy \cite{triginer2020understanding}. In this simplified model, the spontaneously generated mean photon number via $C^{aa}$ is estimated using $\braket{n^c}=\sum_l \sinh^2(c_l E_p)$, where $E_p$ denotes the energy of the injected pump pulse, and $\{c_l\}$ is a set of constants fitted in the low-gain regime. Consequently, the generated mean photon number in the signal mode from the overall system $S$ can be estimated as a reduction from the system $C$, scaled by the escape efficiency of the signal mode (i.e., $\eta_{\text{esc},s}\braket{n^c}$). The scaling factor $\eta_{\text{esc},s}$ arises from the relation between the mean photon numbers in systems $S$ and $C$, as expressed in Eq. (\ref{eq:photon_number_C}).  

When all effects are included, the mean photon number saturates around 0.9, which is considerably lower than the prediction of the simplified SFWM model. Although the quadratic scaling characteristic of the low-gain regime is not explicitly shown in Fig. \ref{fig:fig_nP}(a), the trend indicated by the yellow line falls well below quadratic scaling in the low-gain regime and shows saturation behavior at high-gain regime. This saturation occurs due to shifts in the resonance centers of all three modes, which cause the frequency mismatch at high pump energy, thereby reducing the efficiency of SFWM. The case that includes only SPM, shown by the blue line, demonstrates a slower growth than the simplified SFWM model. The resonance shift in the pump mode caused by SPM reduces the intra-cavity pump energy, thereby impedes the overall effect of SFWM. In contrast, the green line (considering only SFWM) exhibits an exponential increase, even surpassing the scaling of the simplified SFWM model. As discussed earlier, the time-ordering effect enables it to exceed the scaling of the simplified SFWM model. Such faster growth caused by time-ordering correction was also predicted in traveling-wave SPDC \cite{triginer2020understanding}.

The spectral purity ($\mathcal{P}$) and the normalized second-order self-correlation of the signal mode ($g^{(2)}_s$) are plotted in Fig. \ref{fig:fig_nP}(b). The solid lines represent $1+\mathcal{P}$, while the cross-shaped markers denote $g^{(2)}_s$. Previous analytical work has demonstrated that $g^{(2)}_s$ equals $1+\mathcal{P}$ in the lossless case \cite{christ2011probing}. Our results in Fig. \ref{fig:fig_nP}(b) confirm that this equality holds true even in the presence of losses. This suggests that the normalized second-order self-correlation is unaffected by the loss matrices $U_1$ and $U_2$ appearing in the decomposition of Eq. (\ref{eq:sandwich}). Using Eq. (\ref{eq:sandwich}) and Eq. (\ref{eq:unitary_high_Q}), the transfer matrices $S^{aa}_{si}$ and $S^{af}_{si}$ can be expressed in terms of the transfer matrices of the subsystem $C^{aa}$:
\begin{subequations}
\label{eq:S_C_connection}
\begin{align}
\label{eq:S_C_aa_connection}
S^{aa}_{si} &= -\sqrt{\eta_{\text{esc},s}\eta_{\text{esc},i}}C^{aa}_{si},\\
\label{eq:S_C_af_connection}
S^{af}_{si}&= -\sqrt{\eta_{\text{esc},s}(1-\eta_{\text{esc},i})}C^{aa}_{si}.
\end{align}
\end{subequations}
Accordingly, the generated mean photon number in the signal mode, as represented in Eq. (\ref{eq:photon_number}), is given by:
\begin{align}
\label{eq:photon_number_C}
\braket{n_s} &= \eta_{\text{esc},s}\braket{n^c}\nonumber \\&=
\eta_{\text{esc},s}\sum_l \sinh^2(r_l^c). 
\end{align}
Meanwhile, the (unnormalized) second-order self-correlation in Eq. (\ref{eq:self_corr_transfer}) can be simplified as follows:
\begin{align}
\braket{n_s^2} - \braket{n_s} &= \braket{n_s}^2 + \eta_{\text{esc},s}^2\text{Tr}\left[\left(C^{aa}_{si}\left(C^{aa}_{si}\right)^\dagger\right)^2\right] \\&=
\eta_{\text{esc},s}^2 \left[\left(\sum_l\sinh^2\left(r_l^c\right)\right)^2+\sum_l\sinh^4\left(r_l^c\right)\right]. \nonumber
\end{align}
Thus, by canceling the factor $\eta^2_{\text{esc},s}$ from both the numerator and the denominator in the expression for $g^{(2)}_s$, given by Eq. (\ref{eq:norm_self_corr}), it is demonstrated that this quantity remains identical to $1+\mathcal{P}$ even in the presence of loss. 

\begin{figure}[htbp]
\includegraphics[width = \linewidth]{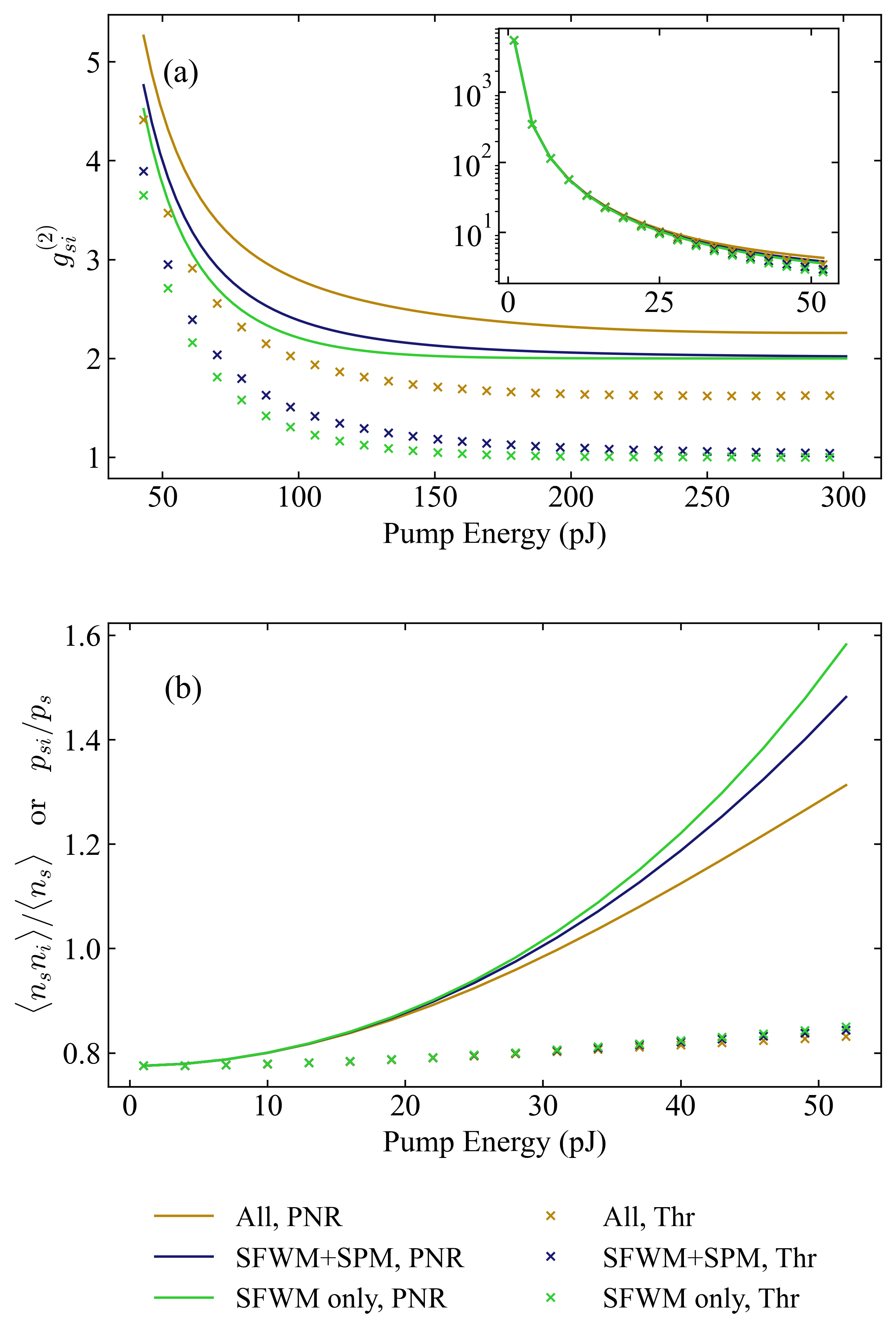}
\caption{(a) Normalized second-order cross-correlation and (b) Unnormalized cross-correlation divided by the signal photon number or the heralding efficiency of the idler photon as functions of pump energy. The solid lines represent results from the PNR detector, while the markers correspond to the threshold detector.}\label{fig:fig_g2}
\end{figure}

Fig. \ref{fig:fig_g2} illustrates the behavior of cross-correlation under various pump energies. The solid lines represent measurements taken with a perfect photon-number resolving (PNR) detector (ensemble averaged), while the dotted lines correspond to measurements with a threshold detector. The detailed formulation for the ensemble average is provided in Appendix \ref{app:sec:physical_quantities}, and the probability of a click event is discussed in Appendix \ref{app:sec:squeezing_parameter}. 

The solid lines in Fig. \ref{fig:fig_g2}(a) show the normalized second-order cross-correlation, $g^{(2)}_{si}$, measured with a PNR detector, with the expression given by Eq. (\ref{eq:norm_cross_corr}). As we can see from the inset, at low pump energies, the lines nearly coincide, regardless of the presence of SPM and XPM. However, as the pump energy increases, all three lines decrease rapidly and saturate at different values. This rapid decrease in $g^{(2)}_{si}$ suggests that the measured coincidence pairs do not arise from the same pair due to multi-photon contributions \cite{takesue2010effects}. The lines without the effect of XPM (green and blue lines) saturate around 2, while the line incorporating the effect of XPM (yellow line) saturates around 2.26. 

This trend in normalized second-order cross-correlation can also be expressed analytically. With the decomposition in Eq. (\ref{eq:sandwich}) and Eq. (\ref{eq:unitary_high_Q}), similar to Eq. (\ref{eq:S_C_connection}), several additional transfer matrices in $S$ can be expressed as follows:
\begin{subequations}
\begin{align}
S^{aa}_{is} &= -\sqrt{\eta_{\text{esc},s}\eta_{\text{esc},i}}C^{aa}_{is}, \\
S^{af}_{is}&= -\sqrt{(1-\eta_{\text{esc},s})\eta_{\text{esc},i}}C^{aa}_{is}, \\
S^{aa}_{ss} &= \eta_{\text{esc},s}C^{aa}_{ss}+(1-\eta_{\text{esc},s})I_N,\\
S^{af}_{ss}&=\sqrt{\eta_{\text{esc},s}(1-\eta_{\text{esc},s})}\left(C^{aa}_{ss}-I_N\right), \\
S^{aa}_{ii} &= \eta_{\text{esc},i}C^{aa}_{ii}+(1-\eta_{\text{esc},i})I_N,\\
S^{af}_{ii} &=\sqrt{\eta_{\text{esc},i}(1-\eta_{\text{esc},i})}\left(C^{aa}_{ii}-I_N\right).
\end{align}
\end{subequations}
Therefore, the (unnormalized) cross-correlation given by Eq. (\ref{eq:cross_corr}) can be simplified as:
\begin{align}
\braket{n_sn_i}=&\braket{n_s}\braket{n_i}\\+&\eta_{\text{esc},s}\eta_{\text{esc},i}\text{Tr}\left[\left(C^{aa}_{si}\right)^*\left(C^{aa}_{ii}\right)^\dagger C^{aa}_{is}\left(C^{aa}_{ss}\right)^T\right]\nonumber\\ =&
\braket{n_s}\braket{n_i}+\eta_{\text{esc},s}\eta_{\text{esc},i}\sum_l \sinh^2(r_l^c)\cosh^2(r_l^c). \nonumber
\end{align}
Consequently, the normalized cross-correlation can be expressed as:
\begin{align}
\label{eq:g2_si_analytic}
g^{(2)}_{si} &= 1 + \cfrac{\sum_l \sinh^2(r_l^c)\cosh^2(r_l^c)}{\left(\sum_l\sinh^2(r_l^c)\right)^2} \nonumber\\ 
&= 1 + \cfrac{1}{\braket{n^c}} + \mathcal{P} \nonumber\\
&= 1 + \cfrac{\eta_{\text{esc},s}}{\braket{n_s}} + \mathcal{P}
\end{align}
The second line in Eq. (\ref{eq:g2_si_analytic}) indicates that the normalized cross-correlation is also unaffected by losses, as the normalized cross-correlation of the subsystem $C^{aa}$ itself is $1+1/\braket{n^c}+\mathcal{P}$ \cite{christ2011probing}. 

The analytic expression in Eq. (\ref{eq:g2_si_analytic}) provides insights into the saturation behavior of $g^{(2)}_{si}$. For the green and blue solid lines, where the effect of XPM is excluded, we observe that the mean photon number increases rapidly and the spectral purity approaches unity with increasing pump energy. Consequently, $g^{(2)}_{si}$ saturates to the value $1+\mathcal{P}\simeq 2$. In contrast, for the yellow line where XPM is included, the mean photon number saturates at approximately 0.9, and the spectral purity becomes about 0.42 when the pump energy reaches $300 \ \mathrm{pJ}$. With the escape efficiency of our system being 0.776, the observed saturation value of 2.26 aligns closely with the analytical estimation provided in Eq. (\ref{eq:g2_si_analytic}).

The dotted lines in Fig. \ref{fig:fig_g2}(a) depict the normalized cross-correlation measured with a threshold detector. The expression for this is given by $p_{si}/p_sp_i$, where $p_{si}$ denotes the probability of a coincidence click event, and $p_{s(i)}$ denotes the probability of a single click event of the signal (idler) photon. At sufficiently low pump energy levels (inset), the dotted lines nearly coincide with the solid lines, regardless of the incorporation of XPM. This is because the mean photon number aligns with the single detection probability, and the (unnormalized) cross-correlation aligns with the coincidence-detection probability. At this level, the quantity $g^{(2)}_{si}$ is equivalent to $1+\mathrm{CAR}$ (Coincidence-to-Accidental ratio), which estimates the noise of a heralded single photon source \cite{signorini2020chip}. The rapid decrease in $g^{(2)}_{si}$ is primarily due to multi-photon contributions, which makes the source less suitable for heralded single photon applications. However, as the pump energy increases, the blue and green dotted lines (without XPM) saturate around 1, while the yellow dotted line (with XPM) saturates around 1.62. When this quantity is saturated around 1, both signal and idler threshold detectors will click with probability near 1 due to very large photon numbers. In this regime, coincidence behavior is completely obscured and no meaningful measurement can be obtained. In contrast, due to the saturation of the mean photon number caused by the incorporation of XPM, the yellow line also saturates, but a value above unity.

Fig. \ref{fig:fig_g2}(b) shows the (unnormalized) cross-correlation divided by the mean photon number of the signal mode (i.e., $\braket{n_sn_i}/\braket{n_s}$), with the solid lines representing this ratio and the dotted lines representing the coincidence probability divided by the detection probability of the signal mode (i.e., $p_{si}/p_s$) by threshold detectors. The pump energies are relatively low in this plot; here, $p_{si}/p_{s}$ is referred to as the heralding efficiency of the idler mode in heralded single photon applications. The quantity reflects the detection probability of the idler photon conditioned on the detection of the signal photon. 

Initially, the values are all close to $0.775$ for both solid and dotted lines, since the ensemble averaged values $\braket{n_sn_i}$ and $\braket{n_s}$ coincide with $p_{si}$ and $p_s$, respectively, at sufficiently low pump energy levels. However, as the pump energy increases, the aligned values deviate, yielding different results between the two measurements. Meanwhile, it is known that the heralding efficiency can be simply estimated by the escape efficiency alone \cite{vernon2016no}. Given that the escape efficiency in our configuration is $0.776$, we confirm from our numerical simulation that the estimation is accurate in the sufficiently low-gain regime. However, the quantity $p_{si}/p_s$ deviates from the estimation value $0.776$ as the pump energy increases. For a threshold detector, the estimation error is less than $7 \%$ up to a pump energy of $50 \ \mathrm{pJ}$, where approximately $0.3$ mean photon numbers are generated.

The plot also reveals that the expression $\braket{n_sn_i}/\braket{n_s}$ is no longer a reliable measure of heralding efficiency as the pump power increases. Even at a moderate pump energy of $25 \ \mathrm{pJ}$, where the differences among the three physical models are negligibly small, mean photon number is approximately $0.08$, however, the values of the two expressions $\braket{n_sn_i}/\braket{n_s}$ and $p_{si}/p_{s}$ are $0.92$ and $0.79$. Note that the pump energy of $25 \ \mathrm{pJ}$ is lower than the energy required to achieve the fundamental limit of single-photon pair generation sources with PNR detectors, where the mean photon number would be 1 \cite{christ2012limits}. Therefore, caution is needed when converting quantum mechanical expressions into measurement probabilities with threshold single photon detectors.

\subsection{Optimal detuning}
The primary cause of spectral distortion in the transfer function and photon number saturation is the non-perturbative effects induced by the combination of SPM and XPM. In the case of CW pumping, it has been shown there exists a detuning that can precisely counteract the resonance shift caused by SPM, thereby preserving the quadratic scaling of the photon pair generation rate \cite{vernon2015strongly}. However, the situation becomes more complex with pulsed pumping. The broadband frequency components of the pump pulse affect SPM differently, leading to the generation of new frequency components from the initial pump pulse. Moreover, the time-dependent behavior of XPM not only shifts the resonance of the signal and idler modes but also distorts their shapes, making the optimal detuning condition for CW operation inapplicable to pulsed operation. 

\begin{figure}[h]
\includegraphics[width = \linewidth]{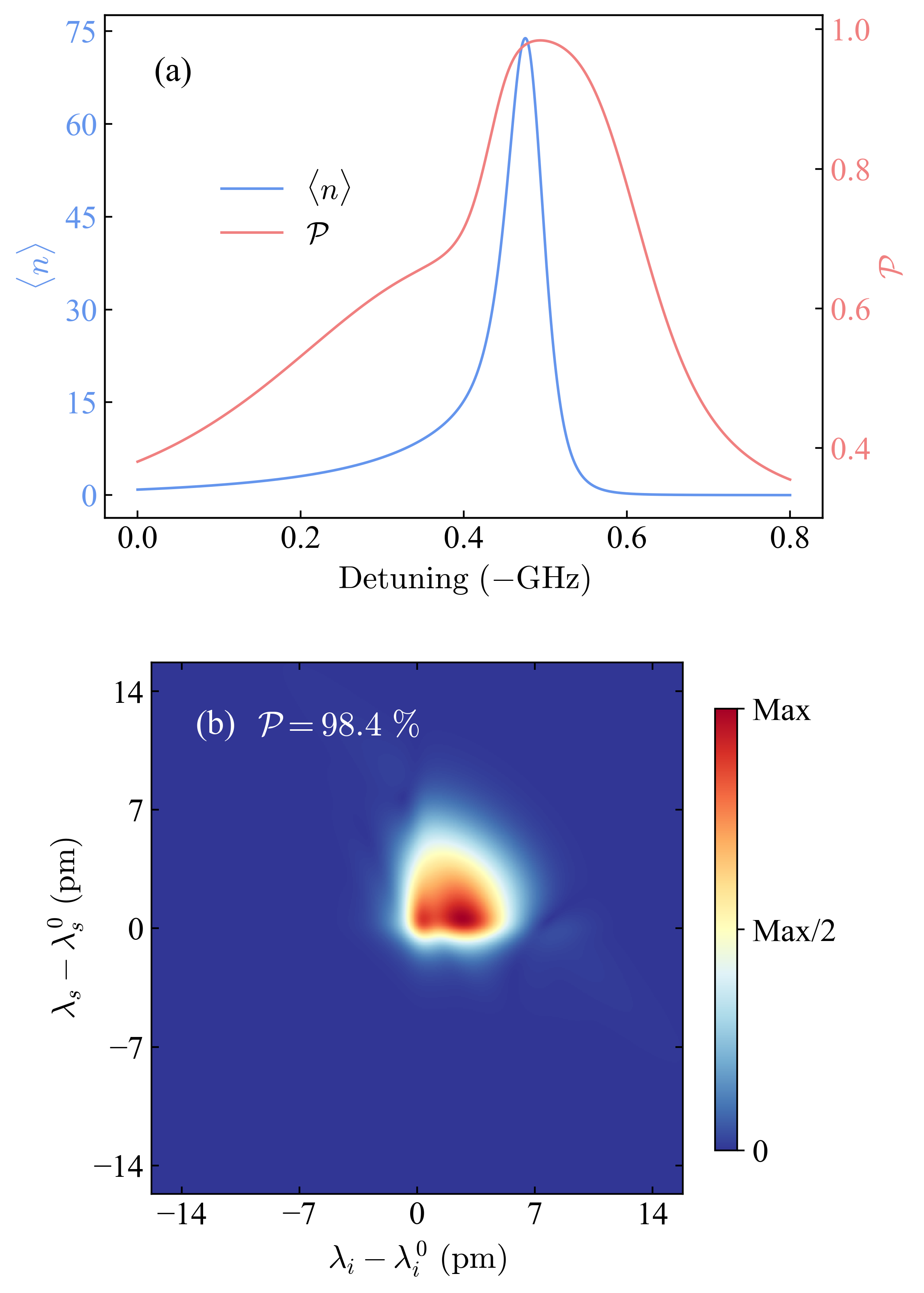}
\caption{(a) Effect of detuning on mean photon number and spectral purity at a pump energy of $600 \ \mathrm{pJ}$. (b) Cross-mode transfer function at maximum spectral purity with a detuning of $-0.49 \ \mathrm{GHz}$.}\label{fig:OD_nPJ}
\end{figure}

In this subsection, we investigate the presence of optimal detuning even with the use of a broadband pump. This approach aims to increase the mean photon number and maximize spectral purity, achieving values exceeding the low-gain bound of approximately $ 93\%$ \cite{vernon2017truly}. We again use the same ring resonator and pump pulse as in the previous section but introduce detuning to the center frequency of the pump pulse. The center frequency is detuned from resonance within the range of $0$ to $-0.8\ \mathrm{GHz}$. With a pump energy fixed at $600 \ \mathrm{pJ}$, the resulting mean photon number and spectral purity are shown in Fig. \ref{fig:OD_nPJ}(a). Both the mean photon number and spectral purity have their own optimal points, termed the optimal detuning for mean photon number and the optimal detuning for spectral purity, respectively. At the optimal detuning for mean photon number, $-0.48 \ \mathrm{GHz}$, the mean photon number reaches approximately $72$. Near the optimal detuning for spectral purity, $-0.49 \ \mathrm{GHz}$, the spectral purity is also maximized around $98.4 \%$, indicating that nearly a single temporal mode is achieved. Fig. \ref{fig:OD_nPJ}(b) displays the cross-mode transfer matrix at this optimal detuning for spectral purity.

\begin{figure}[htbp]
\includegraphics[width = \linewidth]{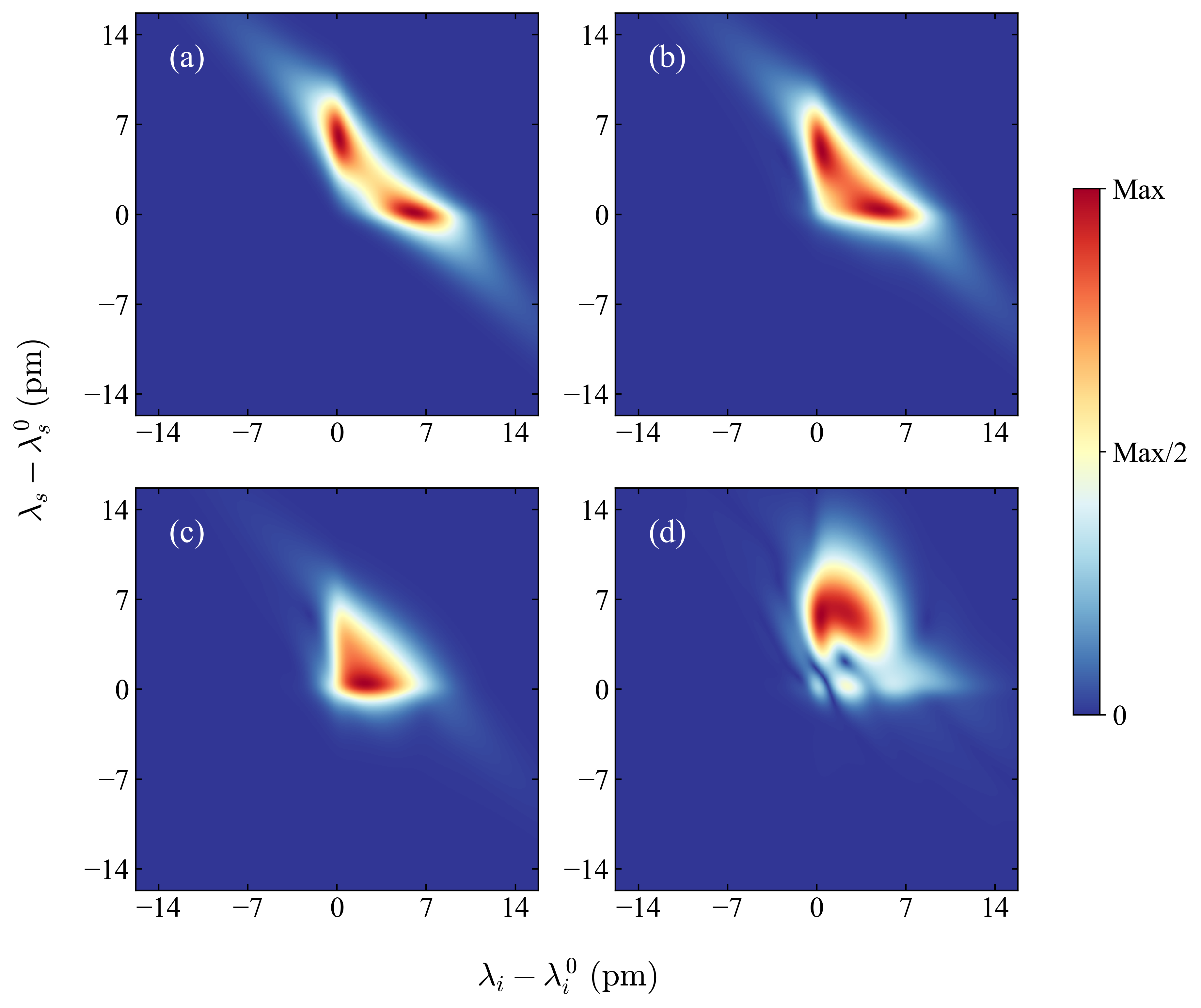}
\caption{Variation of cross-mode transfer functions with increasing pump energy at a fixed detuning of $-0.49 \ \mathrm{GHz}$. The pump energies are: (a) $1 \ \mathrm{pJ}$, (b) $200 \ \mathrm{pJ}$, (c) $400 \ \mathrm{pJ}$, and (d) $800 \ \mathrm{pJ}$.}\label{fig:OD_Ssi_detuned}
\end{figure}

The presence of an optimal point for the maximum mean photon number can be intuitively understood, given the well-established results from classical nonlinear systems \cite{kippenberg2004nonlinear}; however, the reason for the spectral purity also achieving a maximum is less straightforward. To understand this behavior, we first fixed the detuning at $-0.49 \ \mathrm{GHz}$ and varied the pump energy to observe changes in the cross-mode transfer matrix. The cross-mode transfer matrices at pump energies of $1 \ \mathrm{pJ}$, $200 \ \mathrm{pJ}$, $400 \ \mathrm{pJ}$, and $800 \ \mathrm{pJ}$ are shown in Fig. \ref{fig:OD_Ssi_detuned}. 

At a low pump energy of $1 \ \mathrm{pJ}$ in panel (a), we observe a splitting of islands as the detuned pump being frequency mismatched with the resonance frequencies of the signal and idler. This splitting narrows as the pump energy increases and eventually vanishes at a pump energy of $400 \ \mathrm{pJ}$, as shown in panel (b). This effect is primarily caused by XPM, which shifts the resonance frequency and compensates for the frequency mismatch. The cross-mode transfer function shows the merging of distinct islands at higher pump energies, reaching the configuration shown in Fig. \ref{fig:OD_nPJ}(b) at $600 \ \mathrm{pJ}$. At this point, the two islands are merged into one, and the spectral purity is maximized. As the pump energy increases beyond $600 \ \mathrm{pJ}$, the spectral purity decreases due to the more severe effects of XPM distorting the transfer matrix as shown in Fig. \ref{fig:OD_Ssi_detuned}(d). 

This spectral splitting behavior, originated from the interplay between detuning and XPM, is also predicted in the CW case \cite{vernon2015strongly}. However, unlike the CW operation, where the spectrum can be simply expressed as a product of two Lorentzians, the spectrum with a pulsed pump cannot be modeled analytically and results in an unpredictable spectrum. Nonetheless, we observe that as the pump energy increases, the split islands coalesce and form a more rounded mode structure.

\begin{figure}[htbp]
\includegraphics[width = \linewidth]{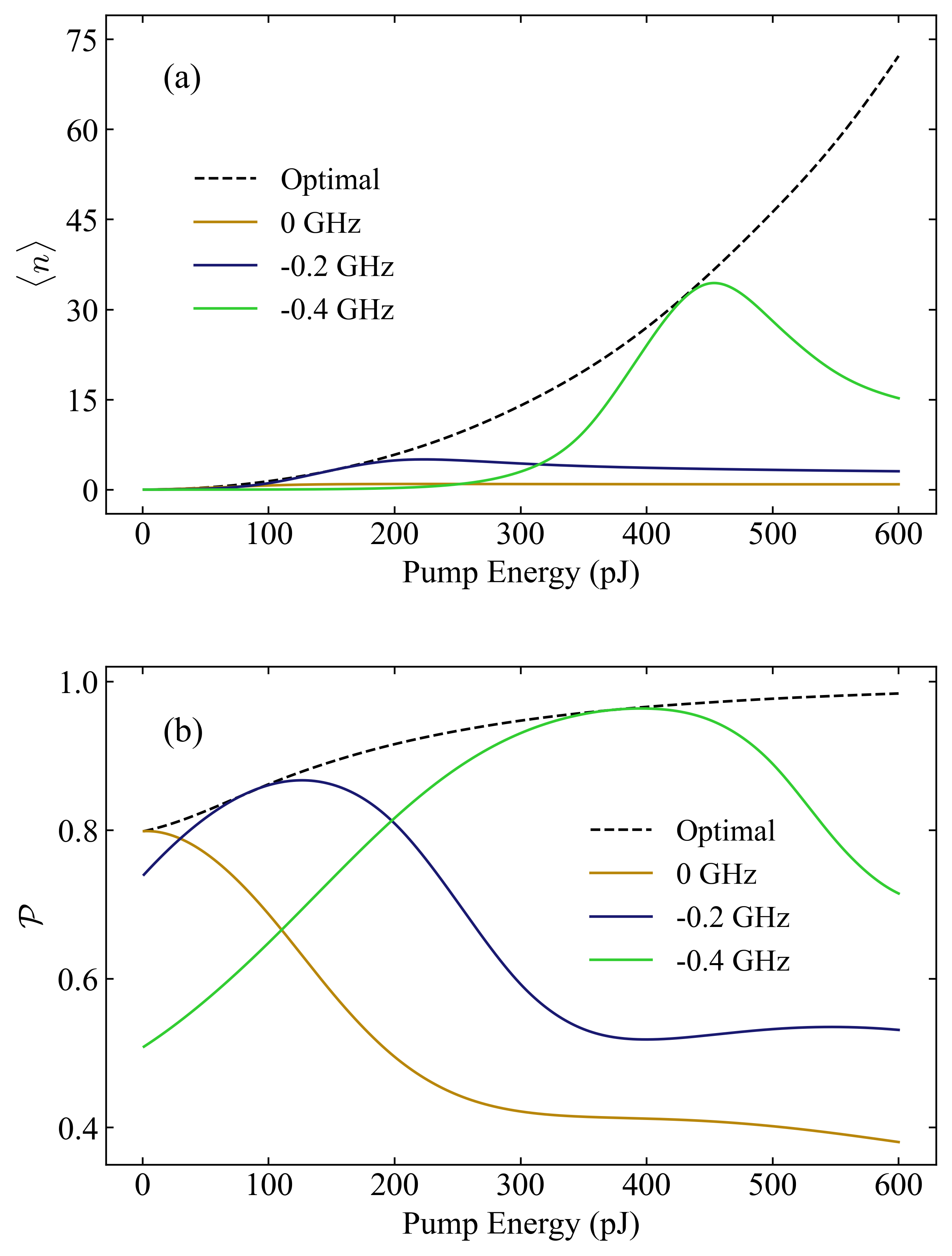}
\caption{Pump energy versus (a) Mean photon number and (b) Spectral purity at various detuning conditions. The black dotted line represents the optimal detuning scenario. The solid yellow, blue, and green lines represent fixed detuning scenarios with $0 \ \mathrm{GHz}$, $-0.2 \ \mathrm{GHz}$, and $-0.4 \ \mathrm{GHz}$, respectively.}\label{fig:OD_optimal_energysweep}
\end{figure}

In general, the optimal detuning for both mean photon number and spectral purity increases monotonically with pump energy. By tracking the optimal detuning at each pump energy, the maximum achievable mean photon number and spectral purity are represented by the dotted line in Fig. \ref{fig:OD_optimal_energysweep}. For comparison, results with fixed detuning are plotted as solid lines. At the optimal points, the mean photon number increases quadratically. As for the spectral purity, it increases monotonically and asymptotically approaches unity when operated with optimal detuning. 

Optimally-detuned pumping consistently outperforms the fixed detuning across all pump energy levels, as shown in Fig. \ref{fig:OD_optimal_energysweep}. The fixed detuning scenarios reach their global maxima for both mean photon number and spectral purity. This behavior is analogous to predictions for photon pair generation rates with CW pumping \cite{vernon2015strongly}. We observe that utilizing optimal detuning in the high-gain regime not only benefits the photon pair generation rate but also enables asymptotic approach to the nearly single temporal mode performance.

\subsection{Two-mode squeezing}
In many high-gain applications, such as CV quantum computation or quantum sensing, a squeezed vacuum serves as a fundamental quantum resource. For these applications, it is essential for the squeezed vacuum to exhibit high levels of squeezing and state purity. This ensures that the variance of the squeezed quadrature is minimized while the state purity, defined by Eq. (\ref{eq:state_purity}), remains high \cite{paris2003purity}. Here, $V_\text{sq}$ and $V_\text{asq}$ represent the variances of the squeezed and anti-squeezed quadratures, respectively, and $\rho$ denotes the density matrix of the state \cite{paris2003purity, vernon2019scalable}. 
\begin{equation}
\label{eq:state_purity}
\text{Tr}\left[\rho^2\right] = \left(V_\text{sq}V_\text{asq}\right)^{-1/2}
\end{equation}
As squeezing strengthens, the state purity typically decreases because the anti-squeezing grows more rapidly than the squeezing due to intrinsic cavity loss. The extent of achievable squeezing is also limited by the cavity's escape efficiency, with its dB representation bounded by $-10\log_{10}(1-\eta_\text{esc})$. Once the squeezing reaches its upper bound, further squeezing cannot be achieved; instead, the anti-squeezing increases the fluctuation of the quadrature. In this situation, further squeezing is prohibited but rather the state purity degrades. Therefore, enhancing the escape efficiency is crucial for generating highly-squeezed and pure quantum light. 

In the previous subsection, it was demonstrated that both spectral purity and mean photon number have their own optimal detunings for maximization. However, for the reasons mentioned above, increasing the mean photon number beyond the squeezing saturation point can degrade the quality of the source. Therefore, in this subsection, we focus on evaluating the application of the ring resonator as a squeezed light source and determine the squeezing parameter using Williamson decomposition, as detailed in Appendix \ref{app:sec:squeezing_parameter} \cite{kopylov2024theory,houde2023waveguided}. We assess the quality of the single-pulsed pump, two-mode squeezed vacuum generated in a single ring resonator across various escape efficiency configurations.

\begin{figure}[htbp]
\includegraphics[width = \linewidth]{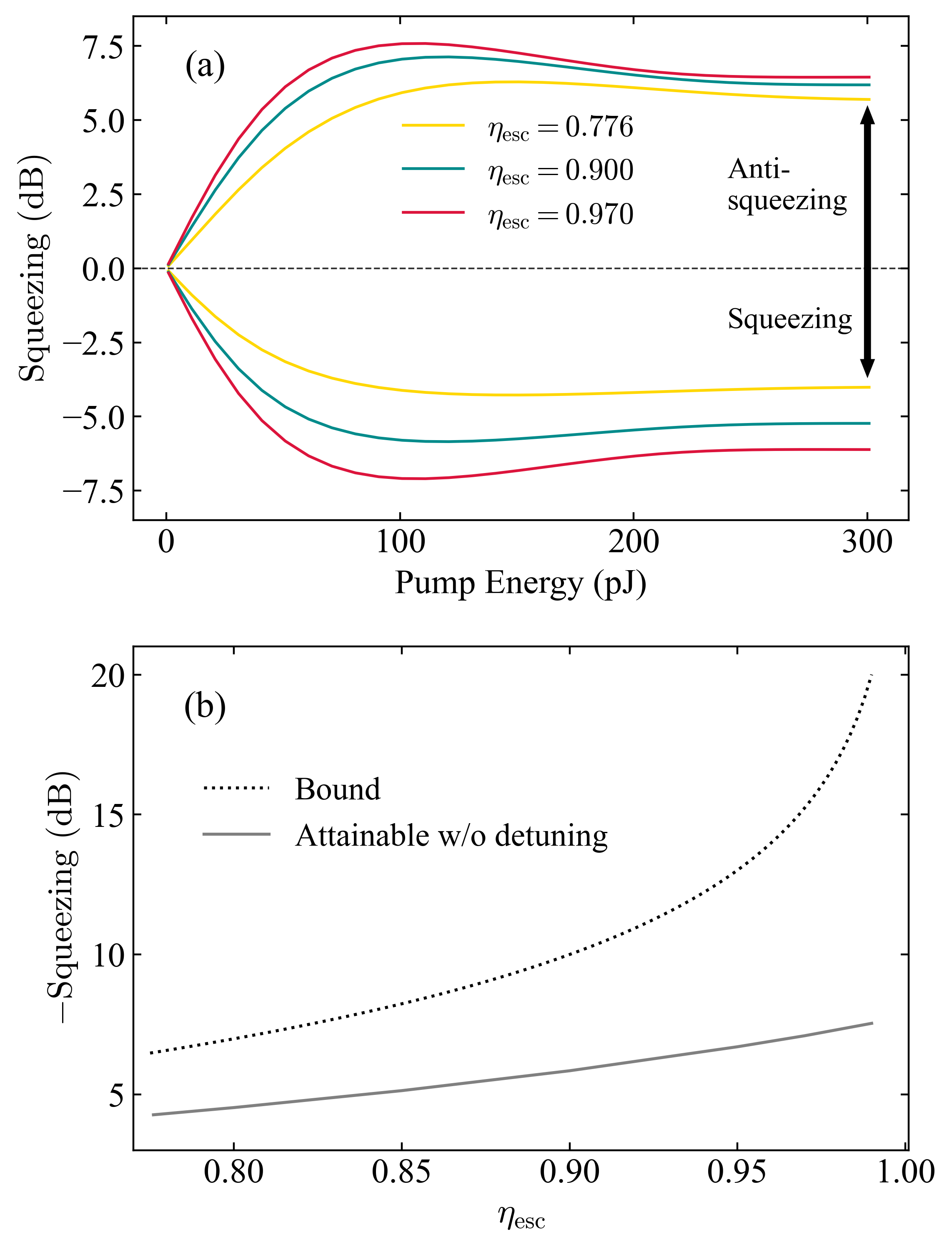}
\caption{(a) Squeezing and anti-squeezing as a function of the pump energy for the various escape efficiencies without detuning. (b) The solid line represents the maximum attainable squeezing given the escape efficiency without detuning. The dotted line represents the upper bound of squeezing determined by the escape efficiency: $-10\log_{10}(1-\eta_\text{esc})$.}\label{fig:SQ_wo_detuning}
\end{figure}

First, we examine the variation in the squeezing parameter without detuning for different escape efficiencies. Fig. \ref{fig:SQ_wo_detuning}(a) illustrates the squeezing and anti-squeezing in dB as functions of pump energy for three escape efficiencies: $0.776$, $0.900$, and $0.970$. These escape efficiencies correspond to propagation losses of $10 \ \mathrm{dB/m}$, $3.86 \ \mathrm{dB/m}$, and $1.08 \ \mathrm{dB/m}$, respectively. A propagation loss of $1.08 \ \mathrm{dB/m}$ is achievable with current state-of-the art single-mode SiN waveguides \cite{liu2022ultralow} and is close to the performance obtainable from the mass-production \cite{alexander2024manufacturable}. We assume that intrinsic loss of the ring is only from the propagation loss, however, scattering from the coupler and the bending losses need to be considered additionally to cover more realistic scenarios. On the figure, negative squeezing represents the squeezed quadrature, while positive squeezing represents the anti-squeezed quadrature for the first Williamson mode. Analogous to the trend of mean photon number shown in Fig. \ref{fig:fig_nP}(a), the squeezing parameter also reaches a global maximum and saturates even with increasing pump energy. Similar to the mean photon number behavior, this saturation is primarily due to XPM which shifts and distorts the photon resonance. XPM moves the resonance away from the frequency-matched point and limits the efficiency of SFWM, thereby limiting the attainable squeezing.

As a result, the maximum attainable squeezing without detuning does not reach the upper bound of $-10\log_{10}(1-\eta_\text{esc})$. To highlight this fact, Fig. \ref{fig:SQ_wo_detuning}(b) compares the maximum attainable squeezing without detuning (solid line) with the upper limit of squeezing determined by escape efficiency (dotted line). The solid line consistently falls below the dotted line, and the difference becomes more pronounced with higher escape efficiencies. Even with an escape efficiency of $0.99$, the maximum attainable squeezing remains below $8 \ \mathrm{dB}$.

\begin{figure}[htbp]
\includegraphics[width = \linewidth]{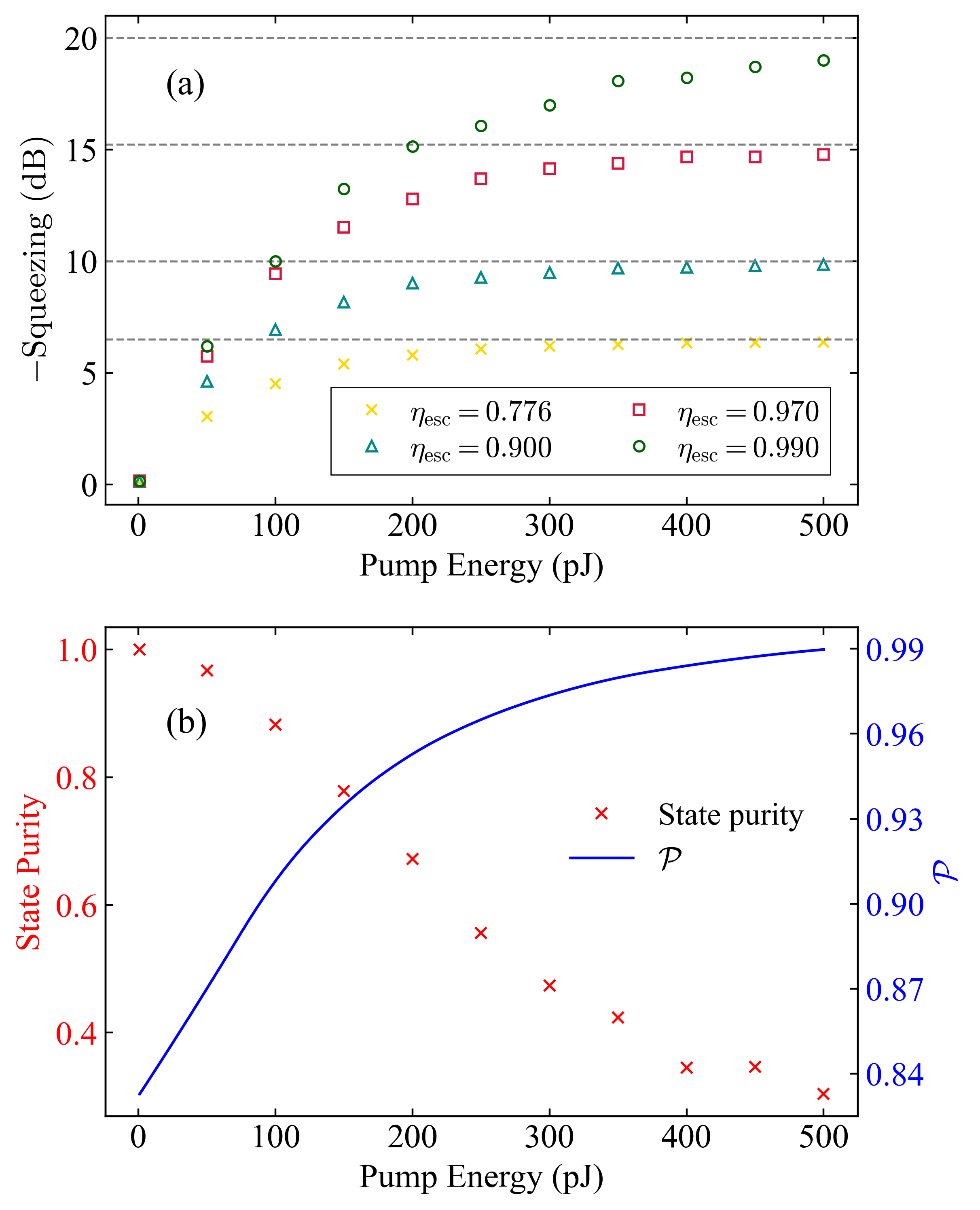}
\caption{(a) Squeezing as a function of the pump energy for the various escape efficiencies at the optimal detuning for spectral purity. (b) State purity and spectral purity versus pump energy at the optimal detuning for spectral purity with an escape efficiency of $0.970$.}\label{fig:SQ_opt_detuning}
\end{figure}

However, detuning the pump pulse effectively enhances both the squeezing and spectral purity simultaneously. At the optimal detuning for spectral purity, we track the squeezing parameters as a function of pump energy. Fig. \ref{fig:SQ_opt_detuning}(a) displays the squeezing versus pump energy for four escape efficiencies: $0.776$, $0.900$, $0.970$, and $0.990$. For each configuration, the upper bounds for squeezing are $6.5 \ \mathrm{dB}$, $10.0 \ \mathrm{dB}$, $15.2 \ \mathrm{dB}$, and $20.0 \ \mathrm{dB}$, respectively, which are indicated by the horizontal grey dotted lines. Across all configurations, the squeezing approaches its upper bound asymptotically as pump energy increases. Detuning enhances the attainable squeezing, enabling it to exceed $8 \ \mathrm{dB}$ even with modest escape efficiencies.

Despite the simultaneous enhancement of squeezing and spectral purity, the state purity deteriorates rapidly as anti-squeezing increases after the squeezing saturates. Fig. \ref{fig:SQ_opt_detuning}(b) illustrates the trade-off between state purity and spectral purity at the optimal detuning for spectral purity with an escape efficiency of $0.970$. When the pump energy is $500 \ \mathrm{pJ}$, causing the squeezing to saturate around $15 \ \mathrm{dB}$, the state purity is approximately $0.3$, while the Schmidt number is $1.01$, indicating near-single temporal mode. Although these values can be optimized by adjusting the ring configuration or pump pulse spectrum, the trade-off is intrinsic; since a decrease in state purity is inevitable with increased squeezing.

\section{CONCLUSION}
\label{sec:discussion}
In this work, we have developed a theoretical model for high-gain, pulsed photon pair generation in a microring resonator. We extended the Heisenberg equations of motion for traveling waves to the resonant case, deriving a broadband transfer matrix that incorporates time-dependent non-perturbative effects. This approach preserves the bosonic commutation relations and is aligned with the TCMT formalism within the high-Q limit. Our formalism can accommodate the low-Q regime, which is relevant for over-coupled cavities, fabrication technologies with moderate propagation loss, or the combination of both.

Our numerical simulations of the transfer functions and associated physical quantities reveal several key insights. Without detuning, the mean photon number saturates and the spectral purity decreases, while the cross-mode transfer function exhibits unpredictable distortions. Notably, the exclusion of XPM leads to a completely inaccurate prediction of the spectrum of the transfer function and photon statistics. In this scenario, the mean photon number does not saturate but scales exponentially, and the spectral purity asymptotically approaches unity. Conversely, the inclusion of XPM significantly distorts the transfer function spectrum and reduces spectral purity as pump energy increases. 

Simulations also address the role of detuning for compensating resonance shifts. Detuning effectively enhances both the mean photon number and spectral purity under high pump energy conditions, allowing for the prediction of optimal points. In some cases, detuning combined with high-gain effects allows the spectral purity to surpass the approximately $93\%$ bound in a single ring resonator operating under low-gain conditions. Optimal detuning and high-energy pumping facilitate quadratic scaling of the mean photon number and an approach toward a single temporal mode. 

The high-gain application in terms of two-mode squeezed vacuum generation is further investigated. Our investigation indicates that squeezing cannot exceed $8 \ \mathrm{dB}$ without detuning under the given device configurations, regardless of escape efficiency. However, optimal detuning can significantly enhance both squeezing and spectral purity simultaneously. Nonetheless, there is an inherent trade-off between spectral purity and state purity, since increased squeezing results in a compromise in state purity.

Our findings contribute to the ongoing research on microring resonator-based photon sources, especially for the CV based researches, whose interests lie in generating highly squeezed states and achieving near single temporal modes. Although experimental efforts have nearly realized these objectives, theoretical support for the full consideration of nonlinear physics under pulsed pump operation has been limited. This paper provides a comprehensive theoretical analysis of the unpredictable behavior of temporal mode structure in the high-gain regime and highlights the benefits of optimal detuning in enhancing squeezing and spectral purity \cite{vaidya2020broadband,arrazola2021quantum}. We anticipate that our results will be instrumental in optimizing and predicting crucial parameters, such as modal structure and photon statistics, for future applications.

\begin{acknowledgments}

This research was supported by the National Research  Foundation of Korea Grant funded by the Korean Government(NRF-2021R1C1C1006400, NRF-2022M3K4A1094782, NRF-2022M3E4A1077013, NRF-2022M3H3A1085772, RS-2024-00442762).

\end{acknowledgments}

\appendix

\section{NONLINEAR COEFFICIENTS}
\label{app:sec:nonlinear_coefficients}
In the main text, we utilized the nonlinear coefficients $\gamma_{\mathrm{sfwm}}$, $\gamma_{\mathrm{spm}}$, and $\gamma_{\mathrm{xpm},s(i)}$. They are determined by the nonlinear susceptibility of the material and the spatial overlap of the modes involved. Their full expressions are as follows:
\begingroup
\allowdisplaybreaks
\begin{widetext}
\begin{subequations}
\begin{align}
\gamma_{\mathrm{sfwm}} &= 3\epsilon_0\sqrt{\bar\omega_s\bar\omega_i}\sum_{ijkl}\int dr_\bot \chi_3^{ijkl}(e_s^i(r_\bot))^*(e_i^j(r_\bot))^*e_p^k(r_\bot)e_p^l(r_\bot), \\
\gamma_{\mathrm{spm}} &= 3\epsilon_0\bar\omega_p\sum_{ijkl}\int dr_\bot \chi_3^{ijkl}(e_p^i(r_\bot))^*(e_p^j(r_\bot))^*e_p^k(r_\bot)e_p^l(r_\bot), \\
\gamma_{\mathrm{xpm},s(i)} &= 3\epsilon_0\bar\omega_{s(i)}\sum_{ijkl}\int dr_\bot \chi_3^{ijkl}(e_{s(i)}^i(r_\bot))^*e_{s(i)}^j(r_\bot)(e_p^k(r_\bot))^*e_p^l(r_\bot).
\end{align}
\end{subequations}
\end{widetext}
\endgroup

\section{LOW-GAIN SOLUTION}
\label{app:sec:low_gain_solution}
In this appendix, we derive a low-gain solution of the transfer functions starting from our formalism, focusing solely on $S^{aa}$ for simplicity. The effects of SPM and XPM are neglected, and the Gaussianity of loss operation is also disregarded since our interest lies in $S^{aa}$. Accordingly, the loss term should be added in Eq. (\ref{eq:photon_pair_dynamics}) to account for the loss, because the phantom waveguide is ignored.

Firstly, neglecting SPM ($\gamma_{\mathrm{spm}}=0$), Eq. (\ref{eq:ikeda_map}) indicates that the intra-cavity field $\beta_p(0,\omega)$ is simply the product of the cavity enhancement function $\mathcal{L}_p(\omega)$ and the incoming field $\beta_p^{\mathrm{in}}(\omega)$: $\beta_p(0,\omega) = \mathcal{L}_p(\omega)\beta_p^{\mathrm{in}}(\omega)$. Here, the cavity enhancement function for mode $m$ is defined as
\begin{equation}
\label{eq:cavity_enhancement}
\mathcal{L}_m(\omega) = \cfrac{i\rho_m}{1-\tau_me^{-\alpha_mL/2}e^{i(\omega-\omega_m^0)T_m}}.
\end{equation}
Thus, the effective pump function Eq. (\ref{eq:B_p}) simplifies to
\begin{align}
\label{eq:B_p_lowgain}
\nonumber B_p(\omega+\omega') = \int d\omega'' &\beta_p^{\mathrm{in}}((\omega+\omega')-\omega'')\beta_p^{\mathrm{in}}(\omega'') \times \\ &\mathcal{L}_p((\omega+\omega')-\omega'')\mathcal{L}_p(\omega''). 
\end{align}

The photon pair generation dynamics, Eq. (\ref{eq:photon_pair_dynamics}), can be analytically solved under the assumptions of perfect matching of both phase and frequency, as well as neglecting XPM. The solution for the signal mode is
\begin{align}
\label{eq:signal_pde_sol_lowgain}
\nonumber
a_s(L,\omega)&e^{-i\Delta k_s(\omega)L}e^{\alpha_sL/2} = a_s(0,\omega) + \\&i\cfrac{\gamma_{\mathrm{sfwm}}L}{2\pi}\int d\omega' B_p(\omega+\omega')a_i^\dagger(0,\omega').
\end{align}
Substituting Eq. (\ref{eq:signal_pde_sol_lowgain}) into the boundary condition Eq. (\ref{eq:photon_bc_simple1}), after removing the phantom waveguide, yields:
\begin{align}
\label{eq:signal_mediate_sol_lowgain}
a_s(0,\omega) &= \mathcal{L}_s(\omega)a_s^{\mathrm{in}}(\omega) +\\
&i\cfrac{\gamma_{\mathrm{sfwm}}L}{2\pi}\left(\cfrac{\mathcal{L}_s(\omega)}{i\rho_s}-1\right)\int d\omega' B_p(\omega+\omega')a_i^\dagger(0,\omega').\nonumber
\end{align}
The input-output relation for the signal mode is obtained by combining Eq. (\ref{eq:signal_mediate_sol_lowgain}) and the second boundary condition Eq. (\ref{eq:photon_bc_simple2}),
\begin{align}
\label{eq:signal_mediate_io_sol_lowgain}
a_s^{\mathrm{out}}(\omega) &= \cfrac{1+i\rho_s\mathcal{L}_s(\omega)}{\tau_s}a_s^{\mathrm{in}}(\omega)+\\
&i\cfrac{\gamma_{\mathrm{sfwm}}L}{2\pi}\cfrac{\mathcal{L}_s(\omega)-i\rho_s}{\tau_s}\int d\omega' B_p(\omega+\omega')a_i^\dagger(0,\omega').\nonumber
\end{align}
Here, the coefficient in the first term $(1+i\rho_s\mathcal{L}_s(\omega))/\tau_s$ represents the transmission function of the signal mode through the cavity. For the compact expression, we define the transmission function of mode $m$ as:
\begin{equation}
\label{eq:transmission}
\mathcal{H}_m(\omega) = \cfrac{\tau_m - e^{-\alpha_mL/2}e^{i(\omega-\omega_m^0)T_m}}{1-\tau_me^{-\alpha_mL/2}e^{i(\omega-\omega_m^0)T_m}}.
\end{equation}
Additionally, $(\mathcal{L}_s(\omega)-i\rho_s)/\tau_s$ in the second term is approximately $\mathcal{L}_s(\omega)$, ignoring a small loss. 

Substituting the idler mode version of Eq. (\ref{eq:signal_mediate_sol_lowgain}) into Eq. (\ref{eq:signal_mediate_io_sol_lowgain}) yields:
\begin{align}
\label{eq:signal_io_lowgain}
a_s^{\mathrm{out}}(\omega) &= \mathcal{H}_s(\omega) a_s^{\mathrm{in}}(\omega) \\ +&i\cfrac{\gamma_{\mathrm{sfwm}}L}{2\pi}\int d\omega' B_p(\omega+\omega')\mathcal{L}_s(\omega)\mathcal{L}_i^*(\omega')\left(a_i^{\mathrm{in}}(\omega')\right)^\dagger \nonumber \\-& \left( \cfrac{\gamma_\mathrm{sfwm}L}{2\pi} \right)^2\iint d\omega' d\omega'' B_p(\omega+\omega')B_p(\omega'+\omega'')\times \nonumber\\ 
&\quad\quad\quad\quad\quad\quad\quad \mathcal{L}_s(\omega)\left(\cfrac{\mathcal{L}_i^*(\omega')}{i\rho_i}+1\right)a_s(0,\omega''). \nonumber
\end{align}
The operator $a_s(0,\omega'')$ that appears in the last term of Eq. (\ref{eq:signal_io_lowgain}) can be further expanded into terms of second-order or higher-order in $\gamma_\mathrm{sfwm}$. Under the first-order perturbation, where terms of second- and higher-orders are all ignored, the transfer functions within the low-gain regime can be calculated as the following:
\begin{subequations}
\label{eq:signal_transfer_functions_lowgain}
\begin{align}
S^{aa}_{ss}(\omega,\omega') &= \mathcal{H}_s(\omega)\delta(\omega-\omega'),\\
\label{eq:signal_transfer_function_si_lowgain}
S^{aa}_{si}(\omega,\omega') &= i\cfrac{\gamma_{\mathrm{sfwm}}L}{2\pi}B_p(\omega+\omega')\mathcal{L}_s(\omega)\mathcal{L}_i^*(\omega'),\\
S^{aa}_{ii}(\omega,\omega') &= \mathcal{H}_i(\omega)\delta(\omega-\omega'),\\
S^{aa}_{is}(\omega,\omega') &= i\cfrac{\gamma_{\mathrm{sfwm}}L}{2\pi}B_p(\omega+\omega')\mathcal{L}_i(\omega)\mathcal{L}_s^*(\omega').
\end{align}
\end{subequations}

\section{PHYSICAL QUANTITIES}
\label{app:sec:physical_quantities}
In this appendix, we extract physical quantities such as mean photon number, second-order correlation, and spectral purity by offering formulations based on transfer matrices from the input-output relation, Eq. (\ref{eq:IO_relation}). 

\medskip

\textit{1. Mean Photon Number}

The mean photon number of the signal mode is given by:
\begin{align}
\label{eq:photon_number}
\nonumber
\braket{n_s^{\mathrm{out}}} &= \int d\omega \braket{(a_s^{\mathrm{out}}(\omega))^\dagger a_s^{\mathrm{out}}(\omega)} \\
&= \text{Tr}\left[S^{aa}_{si}(S^{aa}_{si})^\dagger + S^{af}_{si} (S^{af}_{si})^\dagger\right].
\end{align}
Here, $\text{Tr}[\cdot]$ indicates the trace of the matrix.  

\medskip

\textit{2. Cross-correlation}

The cross-correlation between the signal and idler modes is formulated as:
\begin{align}
\label{eq:coincidence}
&\braket{n_s^{\mathrm{out}}n_i^{\mathrm{out}}} \\
&= \int d\omega d\omega' \braket{(a_s^{\mathrm{out}}(\omega))^\dagger a_s^{\mathrm{out}}(\omega) (a_i^{\mathrm{out}}(\omega'))^\dagger a_i^{\mathrm{out}}(\omega')}. \nonumber
\end{align}
With the continuous input-output relation given by Eq. (\ref{eq:io_conti_form}), the above expression results in eight non-zero ensemble terms:
\begin{subequations}
\begin{align}
&\braket{a_i^\text{in}(\omega_1)\left(a_i^\text{in}(\omega_2)\right)^\dagger a_s^\text{in}(\omega_3)\left(a_s^\text{in}(\omega_4)\right)^\dagger}\\=\nonumber
&\braket{a_i^\text{in}(\omega_1)\left(a_i^\text{in}(\omega_2)\right)^\dagger f_s^\text{in}(\omega_3)\left(f_s^\text{in}(\omega_4)\right)^\dagger}\\=\nonumber
&\braket{f_i^\text{in}(\omega_1)\left(f_i^\text{in}(\omega_2)\right)^\dagger a_s^\text{in}(\omega_3)\left(a_s^\text{in}(\omega_4)\right)^\dagger}\\=\nonumber
&\braket{f_i^\text{in}(\omega_1)\left(f_i^\text{in}(\omega_2)\right)^\dagger f_s^\text{in}(\omega_3)\left(f_s^\text{in}(\omega_4)\right)^\dagger}\\=\nonumber
&\delta(\omega_1-\omega_2)\delta(\omega_3-\omega_4),\\
&\braket{a_i^\text{in}(\omega_1)a_s^\text{in}(\omega_2)\left(a_i^\text{in}(\omega_3)\right)^\dagger\left(a_s^\text{in}(\omega_4)\right)^\dagger}\\=\nonumber
&\braket{a_i^\text{in}(\omega_1)f_s^\text{in}(\omega_2)\left(a_i^\text{in}(\omega_3)\right)^\dagger\left(f_s^\text{in}(\omega_4)\right)^\dagger}\\=\nonumber
&\braket{f_i^\text{in}(\omega_1)a_s^\text{in}(\omega_2)\left(f_i^\text{in}(\omega_3)\right)^\dagger\left(a_s^\text{in}(\omega_4)\right)^\dagger}\\=\nonumber
&\braket{f_i^\text{in}(\omega_1)f_s^\text{in}(\omega_2)\left(f_i^\text{in}(\omega_3)\right)^\dagger\left(f_s^\text{in}(\omega_4)\right)^\dagger}\\=\nonumber
&\delta(\omega_1-\omega_3)\delta(\omega_2-\omega_4).
\end{align}
\end{subequations}
Therefore, the cross-correlation in Eq. (\ref{eq:coincidence}) can be expressed in terms of transfer matrices as:
\begin{align}
\label{eq:cross_corr}
&\braket{n_s^\text{out}}\braket{n_i^\text{out}}+\text{Tr}\Big[\big((S^{aa}_{si})^*(S^{aa}_{ii})^\dagger + (S^{af}_{si})^*(S^{af}_{ii})^\dagger) \times \nonumber \\
&\quad\quad\quad\quad\quad\quad\quad (S^{aa}_{is}(S^{aa}_{ss})^T + S^{af}_{is}(S^{af}_{ss})^T\big)\Big].
\end{align}

Under the low-gain regime, Eq. (\ref{eq:photon_number}) represents the presence probability of a signal photon, while Eq. (\ref{eq:coincidence}) represents the coincidence probability of signal and idler photons. Therefore, their ratio is defined as the heralding efficiency of the idler mode (denoted as $\eta_i$), which is the probability of the presence of idler photon conditioned on the presence of signal photon \cite{signorini2020chip}:
\begin{equation}
\label{eq:h_i}
\eta_i = \cfrac{p_{si}}{p_s}
\simeq
\cfrac{\braket{n_s^{\mathrm{out}}n_i^{\mathrm{out}}}}{\braket{n_s^{\mathrm{out}}}}.
\end{equation}
Also, the normalized cross-correlation $g^{(2)}_{si}$ is defined as:
\begin{equation}
\label{eq:norm_cross_corr}
g^{(2)}_{si} = \cfrac{\braket{n_s^{\mathrm{out}}n_i^{\mathrm{out}}}}{\braket{n_s^{\mathrm{out}}}\braket{n_i^{\mathrm{out}}}}.
\end{equation}

\medskip

\textit{3. Self-correlation}

The self-correlation of the signal mode is:
\begin{align}
\label{eq:self_correlation}
&\braket{(n_s^{\mathrm{out}})^2}-\braket{n_s^{\mathrm{out}}} \\
&=\int d\omega d\omega' \braket{(a_s^{\mathrm{out}}(\omega))^\dagger(a_s^{\mathrm{out}}(\omega'))^\dagger a_s^{\mathrm{out}}(\omega) a_s^{\mathrm{out}}(\omega')} \nonumber.
\end{align}
Using the continuous input-output relation from Eq. (\ref{eq:io_conti_form}), we identify six non-zero ensemble terms:
\begin{subequations}
\begin{align}
&\braket{a_i^\text{in}(\omega_1)a_i^\text{in}(\omega_2)\left(a_i^\text{in}(\omega_3)\right)^\dagger\left(a_i^\text{in}(\omega_4)\right)^\dagger} \\=&\braket{f_i^\text{in}(\omega_1)f_i^\text{in}(\omega_2)\left(f_i^\text{in}(\omega_3)\right)^\dagger\left(f_i^\text{in}(\omega_4)\right)^\dagger} \nonumber\\ =&\delta(\omega_1-\omega_3)\delta(\omega_2-\omega_4) + \delta(\omega_2-\omega_3)\delta(\omega_1-\omega_4), \nonumber  \\
&\braket{a_i^\text{in}(\omega_1)f_i^\text{in}(\omega_2)\left(f_i^\text{in}(\omega_3)\right)^\dagger\left(a_i^\text{in}(\omega_4)\right)^\dagger} \\=&\braket{f_i^\text{in}(\omega_1)a_i^\text{in}(\omega_2)\left(a_i^\text{in}(\omega_3)\right)^\dagger\left(f_i^\text{in}(\omega_4)\right)^\dagger} \nonumber\\ =&\delta(\omega_1-\omega_4)\delta(\omega_2-\omega_3),\nonumber \\
&\braket{a_i^\text{in}(\omega_1)f_i^\text{in}(\omega_2)\left(a_i^\text{in}(\omega_3)\right)^\dagger\left(f_i^\text{in}(\omega_4)\right)^\dagger} \\=&\braket{f_i^\text{in}(\omega_1)a_i^\text{in}(\omega_2)\left(f_i^\text{in}(\omega_3)\right)^\dagger\left(a_i^\text{in}(\omega_4)\right)^\dagger} \nonumber\\ =&\delta(\omega_1-\omega_3)\delta(\omega_2-\omega_4)\nonumber.
\end{align}
\end{subequations}
Using the results above, the self-correlation in Eq. (\ref{eq:self_correlation}) can be expressed in terms of the transfer matrices as:
\begin{equation}
\label{eq:self_corr_transfer}
\braket{n_s^\text{out}}^2 + \text{Tr}\left[\left(S^{aa}_{si}(S^{aa}_{si})^\dagger + S^{af}_{si}(S^{af}_{si})^\dagger\right)^2\right].
\end{equation}

The normalized self-correlation $g^{(2)}_{s}$ is defined as:
\begin{equation}
\label{eq:norm_self_corr}
g^{(2)}_{s} = \cfrac{\braket{(n_s^{\mathrm{out}})^2}-\braket{n_s^{\mathrm{out}}}}{\braket{n_s^{\mathrm{out}}}^2},
\end{equation}
which is called the unheralded $g^{(2)}$ of the signal mode.

\medskip

\textit{4. Schmidt Number and Spectral Purity}

The Schmidt number $\mathcal{K}$ estimates the effective number of emitted Schmidt modes, and the spectral purity $\mathcal{P}$ is its reciprocal. The broadband Schmidt modes can be extracted from the decomposition of $C^{aa}$, as defined in Eq. (\ref{eq:C_aa_decomp}). The Schmidt number and spectral purity are given by the following expression:
\begin{equation}
\mathcal{K} = 1/\mathcal{P}=\cfrac{\left(\sum_l \sinh^2{\left(r_l^c\right)}\right)^2}{\sum_l \sinh^4{\left(r_l^c\right)}},
\end{equation}
where $\{r_l^c\}$ are the diagonal components of $R^c$. The diagonal components represent the squeezing parameters of the Schmidt mode $l$ \cite{triginer2020understanding} for the lossless system $C^{aa}$. 

Additionally, the spectral purity can also be extracted from the Schmidt decomposition of the cross-mode transfer matrix $S^{aa}_{si}$, since it is a simply amplitude modified version of $C^{aa}_{si}$, as expressed by Eq. (\ref{eq:S_C_aa_connection}).

These formulations provide clear expressions of the key system metrics such as Schmidt numbers, photon statistics and self- and cross-correlations, which can be derived from the transfer matrices.

\section{GAUSSIAN QUANTUM OPTICS FRAMEWORK}
\label{app:sec:squeezing_parameter}
The input-output relation in Eq. (\ref{eq:IO_relation}) encapsulates the overall information of the system's Gaussian unitary transformation. In this appendix, we employ the frameworks of the Gaussian quantum optics and illustrate how the threshold detector is incorporated and the squeezing parameter is obtained \cite{horoshko2019bloch,thomas2021general}. 

\medskip

\begingroup
\allowdisplaybreaks
\begin{figure*}[htbp]
\centering
\includegraphics[width = 0.9\textwidth]{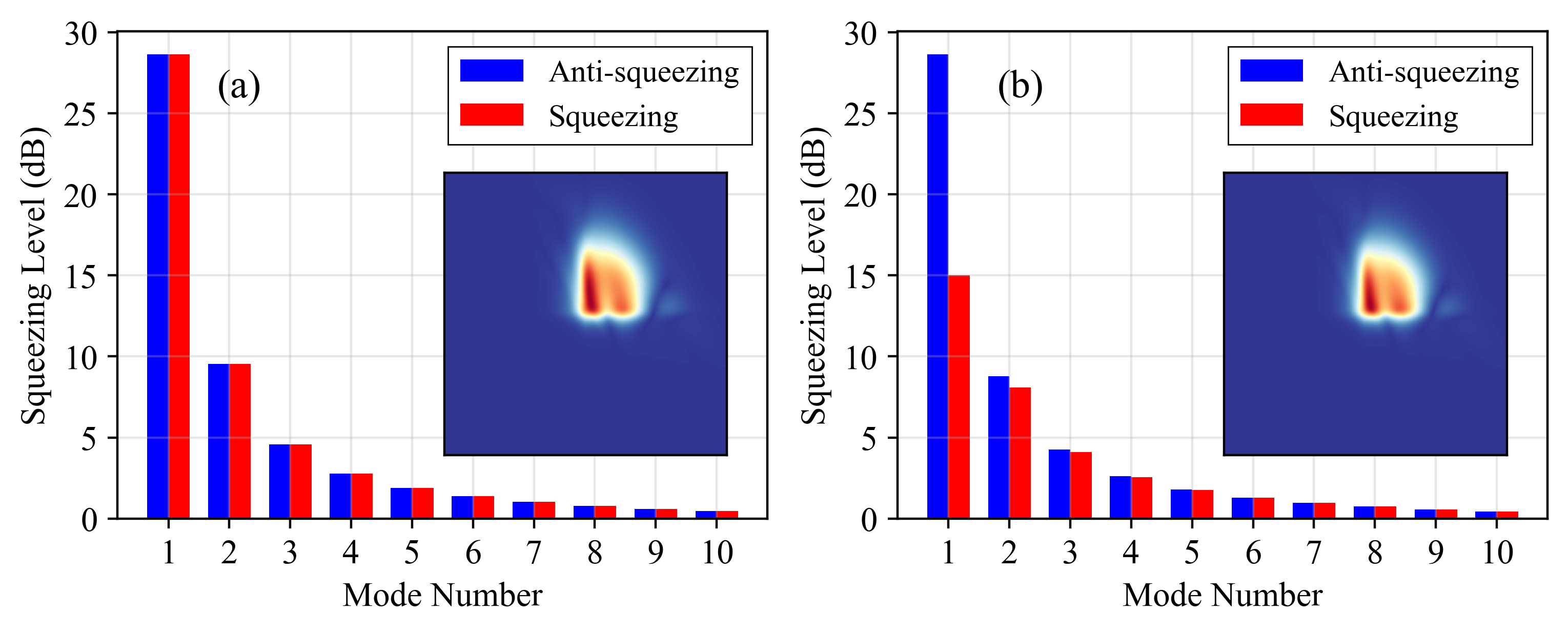}
\caption{Anti-squeezing (blue bar) and squeezing (red bar) at escape efficiencies of: (a) $\eta_\text{esc}=1$, (b) $\eta_\text{esc}=0.97$. The inset shows the corresponding cross-mode transfer function $S^{aa}_{si}$.}\label{fig:app_wil}
\end{figure*}
\endgroup

\textit{1. Complex Symplectic Matrix and Covariance Matrix}

The complex symplectic matrix is a linear transfer matrix of an operator vector $[\vec a_s, \vec a_i, \vec f_s, \vec f_i, \vec a_s^\dagger, \vec a_i^\dagger, \vec f_s^\dagger, \vec f_i^\dagger]^T$. The complex symplectic matrix $M$ can be expressed in terms of block matrices of $S$, as shown in Eq. (\ref{eq:complex_symplectic}). Each block matrix is of size $N\times N$, where $N$ denotes the number of frequency discretizations.
\begingroup
\allowdisplaybreaks
\begin{widetext}
\begin{equation}
\label{eq:complex_symplectic}
M = \begin{pmatrix}
S^{aa}_{ss} & O_N & S^{af}_{ss} & O_N & O_N & S^{aa}_{si} & O_N & S^{af}_{si} \\
O_N & S^{aa}_{ii} & O_N & S^{af}_{ii} & S^{aa}_{is} & O_N & S^{af}_{is} & O_N \\
S^{fa}_{ss} & O_N & S^{ff}_{ss} & O_N & O_N & S^{fa}_{si} & O_N & S^{ff}_{si} \\
O_N & S^{fa}_{ii} & O_N & S^{ff}_{ii} & S^{fa}_{is} & O_N & S^{ff}_{is} & O_N \\
O_N & (S^{aa}_{si})^* & O_N & (S^{af}_{si})^* & (S^{aa}_{ss})^* & O_N & (S^{af}_{ss})^* & O_N \\
(S^{aa}_{is})^* & O_N & (S^{af}_{is})^* & O_N & O_N & (S^{aa}_{ii})^* & O_N & (S^{af}_{ii})^* \\
O_N & (S^{fa}_{si})^* & O_N & (S^{ff}_{si})^* & (S^{fa}_{ss})^* & O_N & (S^{ff}_{ss})^* & O_N \\
(S^{fa}_{is})^* & O_N & (S^{ff}_{is})^* & O_N & O_N & (S^{fa}_{ii})^* & O_N & (S^{ff}_{ii})^*
\end{pmatrix}.
\end{equation}
\end{widetext}
\endgroup
The matrix $M$ satisfies a symplectic condition $M\mathbb{K}M^\dagger = \mathbb{K}$ to preserve the bosonic commutation relations, where the matrix $\mathbb{K}$ is defined as Eq. (\ref{eq:K}).
\begin{equation}
\label{eq:K}
\mathbb{K} = \begin{pmatrix} I_{4N} & 0_{4N} \\ 0_{4N} & -I_{4N}\end{pmatrix}.
\end{equation}

The transformation described by $M$ is a Gaussian process, which transforms one Gaussian state into another. A Gaussian state is uniquely characterized by its complex covariance matrix $\sigma$. Without considering a displacement in phase space, the state's complex covariance matrix evolves as shown in Eq. (\ref{eq:cov_evolve}).
\begin{equation}
\label{eq:cov_evolve}
\sigma^{\mathrm{out}} = M\sigma^{\mathrm{{in}}}M^\dagger.
\end{equation}
Assuming the initial state is a vacuum state with a complex covariance matrix $\sigma^{\mathrm{in}} = I_{8N}$, the output complex covariance matrix is $\sigma^\mathrm{out} = MM^\dagger$. The complex covariance matrix is Hermitian and positive-definite.

\medskip

\textit{2. Threshold Detector}

Unlike the ensemble average calculations detailed in Appendix \ref{app:sec:physical_quantities}, the Gaussian quantum optics formalism allows us to calculate the probability distribution of the state by employing detectors. A threshold detector enables binary classification: detection (click) or no detection (non-click), whereas a PNR detector provides the probability distribution of the number of detected photons. Here, we focus on the threshold detector and describe how to obtain the probabilities of click and non-click events using the covariance matrix. 

The Gaussian nature of the state allows us to trace out the unwanted modes and focus solely on the modes of interest by removing the rows and columns corresponding to these unwanted modes from the covariance matrix. From the entire set of modes $\{\text{signal}, \text{idler}, \text{ph-signal}, \text{ph-idler}\}$, where $\text{ph-}$ denotes the modes from the phantom waveguide, we classify the set of modes of interest as $\bb{S}$, and the modes to be traced out as $\bb{L}$. For instance, to calculate the detection probability of the signal mode, we set $\bb{S} = \{ \text{signal} \}$ and $\bb{L} = \{\text{idler}, \text{ph-signal}, \text{ph-idler}\}$. Similarly, to calculate the coincidence detection probability of the signal and idler modes, we set $\bb{S} = \{\text{signal}, \text{idler}\}$ and $\bb{L} = \{\text{ph-signal}, \text{ph-idler}\}$.

Let $\sigma_{\bb{S}}$ be the covariance matrix after tracing out the unwanted modes. The probability of projecting onto the vacuum state yields the probability of a non-click event, given by \cite{kim2024simulation}:
\begin{equation}
\label{eq:P_off}
P_\mathrm{off}(\bb{S}) = \mathrm{Tr}\left[\rho\ket{\text{vac}}\bra{\text{vac}}_\bb{S}\right] = \left(\mathrm{det}\left[(I_{2|\bb{S}|}+\sigma_\bb{S})/2\right]\right)^{-1/2},
\end{equation}
where $|\bb{S}|$ represents the number of modes in $\bb{S}$. Since threshold detection results in binary outcomes, the probability of a click event is:
\begin{equation}
\label{eq:P_on}
P_\text{on}(\bb{S}) = 1 - P_\text{off}(\bb{S}).
\end{equation}
In a same way, the probability of coincidence in modes $\bb{S}$ and $\bb{S}'$ is:
\begin{equation}
P_\text{coin}(\bb{S}, \bb{S}') = 1-P_\text{off}(\bb{S})-P_\text{off}(\bb{S'}) + P_\text{off}(\bb{S}, \bb{S}').
\end{equation}
Thus, the detection probability for the single mode, $p_{s(i)}$, and the coincidence probability, $p_{si}$, are estimated as described in the main text.

\medskip

\textit{3. Squeezing Parameter}

The squeezing parameters for multi-mode Gaussian transformations are determined using a specific broadband basis. In the main text, the squeezing parameters are expressed in the Williamson basis. The Williamson basis, along with its associated squeezing parameters, is derived through the Williamson decomposition and the Bloch-Messiah (or Euler) decomposition applied to the real covariance matrix. 

The real covariance matrix represents the covariance of the quadrature operators. The set of photon creation and annihilation operators $[\vec a_s, \vec a_i, \vec a_s^\dagger, \vec a_i^\dagger]^T$ is transformed into the set of quadrature operators $[\vec x_s, \vec x_i, \vec p_s, \vec p_i]^T$, where $x_j$ and $p_j$ are defined as:
\begin{equation}
x_j = \cfrac{a_j + a_j^\dagger}{\sqrt{2}}, \quad p_j = \cfrac{a_j - a_j^\dagger}{\sqrt{2}i}.
\end{equation}
Accordingly, the vector transformation matrix $\mathbb{R}$ is defined as:
\begin{align}
 \begin{pmatrix}\vec x_s\\ \vec x_i\\ \vec p_s\\  \vec p_i\end{pmatrix} = &\mathbb{R}
\begin{pmatrix}\vec a_s\\ \vec a_i\\ \vec a_s^\dagger\\ \vec a_i^\dagger\end{pmatrix}\\
\text{where} \ \ \mathbb{R} =\cfrac{1}{\sqrt{2}}& \begin{pmatrix} I_{2N}& I_{2N} \\ -iI_{2N}& iI_{2N} \end{pmatrix}.\nonumber
\end{align}
Consequently, the complex covariance matrix of the subspace $\bb{S} = \{\text{signal, idler}\}$, $\sigma_\bb{S}$, is transformed into the real covariance matrix $V_\bb{S}$ via the basis transformation:
\begin{equation}
V_\bb{S} = \mathbb{R}\sigma_\bb{S}\mathbb{R}^\dagger.
\end{equation}
The real covariance matrix $V_\bb{S}$ is positive-definite and real symmetric. 

The Williamson decomposition breaks down any positive definite real matrix into a real symplectic matrix $S_\text{w}$ and diagonal matrix $D_\text{th}$ with real and positive values. This decomposition is expressed as:
\begin{align}
V_\bb{S} &= S_\text{w} D_\text{th} S_\text{w}^T \\
\text{where}\ \ D_\text{th} = \text{diag}&\left(\nu_1, ..., \nu_{2N}, \nu_1, ..., \nu_{2N}\right)\nonumber.
\end{align}
Here, the components $\{\nu_j\}$ are known as the symplectic eigenvalues of the real covariance matrix. The diagonal components of $D_\text{th}$ represent an equivalent system initialized with each mode in a thermal state. In the absence of internal cavity loss, all symplectic eigenvalues are 1, indicating that the states are in a pure vacuum state. Physically, the Williamson decomposition decouples thermal fluctuation in $D_\text{th}$ from the squeezing process (quantum fluctuation) in $S_\text{w}$. 

The Bloch-Messiah decomposition is applicable to any real symplectic matrix, breaking it down into two orthogonal symplectic matrices $O_l$ and $O_r$, and a diagonal matrix $\Lambda$. Applying the Bloch-Messiah decomposition to $S_\text{w}$ yields:
\begin{align}
S_\text{w} &= O_l\Lambda O_r \\
\text{where} \ \ \Lambda = \text{diag}(e^{r_1}, &e^{r_1}, ...,e^{r_N}, e^{r_N},\nonumber\\
& \quad  e^{-r_1}, e^{-r_1}, ..., e^{-r_N}, e^{-r_N}).\nonumber
\end{align}
The multiplicity of 2 in the diagonal components of $\Lambda$ arises because $S_\text{w}$ pertains to two-mode squeezing. From a physical perspective, the Bloch-Messiah decomposition constructs an equivalent system where two passive Gaussian transformations $O_l$ and $O_r$ sandwich a multiple independent single-mode squeezers $\Lambda$, characterized by the squeezing parameters $(r_1,...,r_N)$. 

By combining the Williamson decomposition and the Bloch-Messiah decomposition, the real covariance matrix $V_\bb{S}$ is decomposed as follows:
\begin{equation}
V_\bb{S} = O_l \Lambda O_r D_\text{th} O_r^T \Lambda O_l^T.
\end{equation}
The Williamson mode basis $U_\text{w}$ is defined from the symplectic orthogonal matrix $O_l$, expressed as:
\begin{equation}
O_l  = \begin{pmatrix} \text{Re}(U_\text{w}) &  -\text{Im}(U_\text{w}) \\ \text{Im}(U_\text{w}) & \text{Re}(U_\text{w})\end{pmatrix}.
\end{equation}
In this basis, the diagonalized real covariance matrix is:
\begin{equation}
\Sigma_\text{w} = \Lambda O_r D_\text{th} O_r^T \Lambda.
\end{equation}
Among the diagonal components of $\Sigma_\text{w}$, the maximum value $\Sigma_\text{w}^\text{max}$ is associated with the variance of the anti-squeezed quadrature of the first Williamson mode, while the minimum value $\Sigma_\text{w}^\text{min}$ is associated with the variance of the squeezed quadrature of the first Williamson mode. The squeezing and anti-squeezing of the first Williamson mode in dB are expressed as:
\begin{subequations}
\begin{align}
\text{Squeezing (dB)} = -10\log_{10}\left(\Sigma_\text{w}^\text{min}\right),& \\
\text{Anti-squeezing (dB)} = 10\log_{10}\left(\Sigma_\text{w}^\text{max}\right)&.
\end{align}
\end{subequations}

Fig. \ref{fig:app_wil} shows the squeezing and anti-squeezing values of the first ten Williamson modes (ignoring multiplicity) at escape efficiencies of $1$ and $0.97$, respectively. As expected, while the squeezing and anti-squeezing coincide in the lossless case shown in panel (a), in the lossy case of panel (b), the squeezing of the first Williamson mode is limited to around $15 \ \mathrm{dB}$. Therefore, the squeezing of the first Williamson mode degrades the most, while higher-order modes are less affected.

\section{Multiple Phantom Channels}
\label{app:sec:multiple_phantom_channels}

\begin{figure}[htbp]
    \centering
    \includegraphics[width=0.62\linewidth]{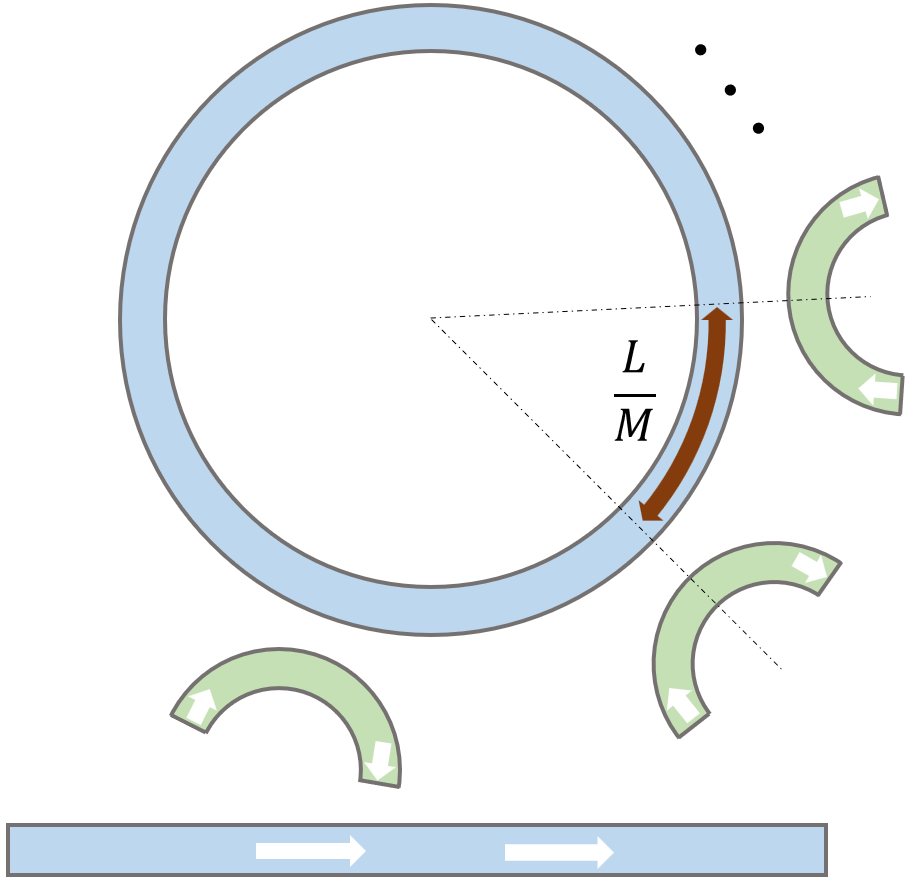}
    \caption{Illustration of a ring resonator coupled to $M$ phantom channels, equally spaced by $L/M$.}
    \label{fig:cav_mulphan}
\end{figure}

In the main text, we introduced our frequency-domain simulation framework using a single phantom channel to account for internal loss. This approach simplifies the model, while keeping an essential functionality for exploring underlying physics. In the linear domain, without nonlinearity, a transfer function approach with a single phantom channel is sufficient for an exact description of Airy-shaped resonances, in contrast to the Lorentzian profiles described by TCMT. However, when combined with nonlinearities, the use of a single phantom channel imposes limitations on accurately capturing the effects of loss in low-finesse regimes \cite{banic2022two}. For a more precise representation of internal loss, multiple phantom channels must be incorporated \cite{sloan2024high, kim2024simulation}, as illustrated in Fig. \ref{fig:cav_mulphan}, to better address the propagation loss of quantized fields within the resonator. In this appendix, we analyze the impact of the number of phantom channels on simulation accuracy. 

The multiple phantom channel model can be integrated into our transfer matrix framework by introducing multiple boundary conditions. We consider $M$ equally spaced phantom channels, represented by the green waveguides in Fig.\ref{fig:cav_mulphan}. For each phantom waveguide, distinct noise operators $f^{\text{in},(l)}$, indexed by $l=1\sim M$, couple to the ring resonator. The transmission and reflection coefficients for each channel are given by:
\begin{subequations}
    \begin{align}
        \tilde\gamma &= \exp\left[-\cfrac{\alpha L}{2M}\right],\\
        \tilde\kappa &= \sqrt{1-\tilde\gamma^2},
    \end{align}
\end{subequations}
respectively. 

\begin{figure}[htbp]
    \centering
    \includegraphics[width=\linewidth]{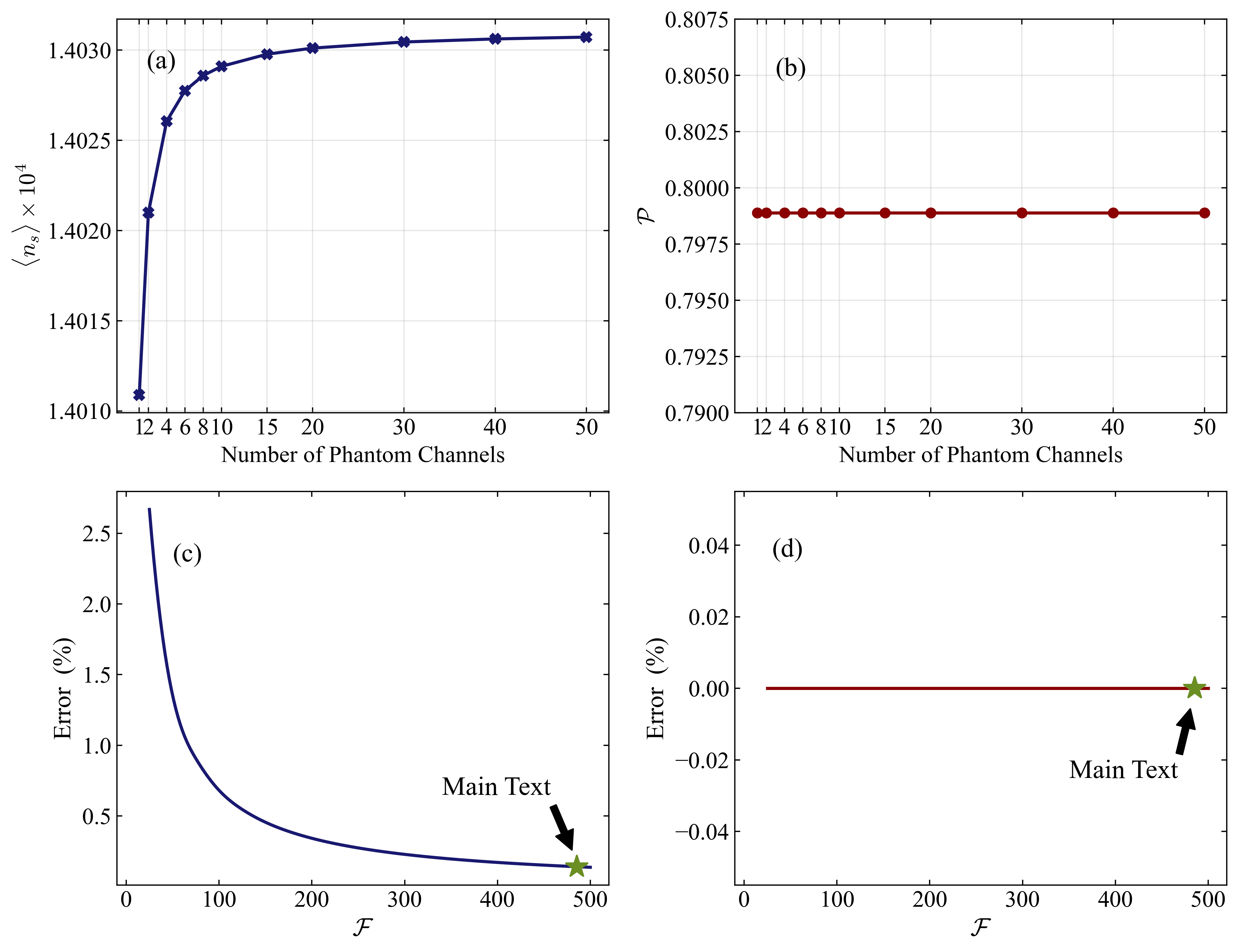}
    \caption{(a, b) Effect of the number of phantom channels on the mean photon number and spectral purity. (c, d) Simulation error of a single phantom channel model relative to the converged value as a function of finesse ($\mathcal F$), with a fixed escape efficiency of 0.776. The star marker indicates $\mathcal F = 485$, the configuration used in the main text.}
    \label{fig:mulphan_sim}
\end{figure}

Using the same configuration of the ring resonator and pump pulse as in the main text (see Sec. \ref{sec:numerical_simulation}), we evaluated the mean photon number and spectral purity as functions of the number of phantom channels, as shown in panel (a) and (b) of Fig. \ref{fig:mulphan_sim}. The results indicate that the mean photon number converges as the number of phantom channels increases, while the spectral purity remains constant. This demonstrates that the single phantom channel model is sufficient for accurately simulating spectral purity and corresponding temporal mode structure. Furthermore, the mean photon number calculated with a single phantom channel shows an error of less than $0.2\%$ compared to the converged value. Given that the simulations in the main text were conducted in a high-finesse regime (finesse of 485), the single phantom channel model exhibits minimal discrepancies compared to the multiple phantom channel model. This behavior persists even under strong nonlinearity conditions, such as high pump energy or optimal detuning, while the results in Fig. \ref{fig:mulphan_sim} was obtained in the low pump energy of $1 \ \mathrm{pJ}$. 

We further evaluated the error in the mean photon number and spectral purity of the single phantom channel model as functions of finesse, while maintaining a fixed escape efficiency of 0.776. The error was determined relative to the converged values obtained using multiple phantom channels. The results are presented in panel (c) and (d) of Fig. \ref{fig:mulphan_sim}, where we marked the finesse value of 485 used in the main text (indicated by a star). As finesse decreases, the error in the mean photon number increases; however, it remains below $1\%$ for finesse values greater than 70. Moreover, spectral purity remains unchanged even in the low-finesse regime. Nonetheless, the increasing error in the mean photon number suggests that the multiple phantom channel model should be employed for systems operating in the low-finesse regime, approximately below a finesse value of 50.

\bibliographystyle{unsrt}

\bibliography{biblio}

\end{document}